\documentclass[12pt,nohyper]{JHEP}

\usepackage{amsmath,amssymb,amscd,subfigure}
\usepackage[dvips]{graphicx}

\newcommand{\pI}{\mathsf{p1}} 
\newcommand{\pg}{\mathsf{pg}}
\newcommand{\pgg}{\mathsf{pgg}}
\newcommand{\pII}{\mathsf{p2}}
\newcommand{\pM}{\mathsf{pm}}
\newcommand{\cm}{\mathsf{cm}}
\newcommand{\pmm}{\mathsf{pmm}}
\newcommand{\cmm}{\mathsf{cmm}}
\newcommand{\pmg}{\mathsf{pmg}}
\newcommand{\pIII}{\mathsf{p3}}
\newcommand{\pIIImI}{\mathsf{p3m1}}
\newcommand{\pIIIIm}{\mathsf{p31m}}
\newcommand{\pIV}{\mathsf{p4}}
\newcommand{\pIVm}{\mathsf{p4m}}
\newcommand{\pIVg}{\mathsf{p4g}}
\newcommand{\pVI}{\mathsf{p6}}
\newcommand{\pVIm}{\mathsf{p6m}}

\newcommand {\Zb} {\mathbb{Z}} 
\newcommand {\Qb} {\mathbb{Q}} 
\newcommand {\Rb} {\mathbb{R}} 
\newcommand {\Cb} {\mathbb{C}} 
\newcommand {\Nb} {\mathbb{N}} 
\newcommand {\Db} {\mathbb{D}} 
\newcommand {\Fb} {\mathbb{F}}
\newcommand {\pz} {\partial_z}
\newcommand {\pzc} {\partial_{z^*}}
\newcommand {\dzz} {\frac{i \, dz \, dz^*}{2(\operatorname{Im} \omega)^2}}

\newcommand {\Lag} {\mathcal{L}}
\newcommand {\I} {\operatorname{I}}
\newcommand {\II} {\operatorname{II}}
\newcommand {\III} {\operatorname{III}}
\newcommand {\id} {\mathbf{1}} 
\newcommand {\vp} {\varphi}
\newcommand {\rct} {\tilde{r}_c}
\newcommand {\GF} {\operatorname{GF}}
\newcommand {\FP} {\operatorname{FP}}
\newcommand{\dBKT}{\bigl[ dz \, dz^* \bigr]_{\operatorname{BKT}}}

\newcommand{\Orbifold}[2]{
\subfigure[#1]{
  \begin{minipage}[t]{0.45\textwidth}
    \parbox[b]{\textwidth}{
      \centering
      \scalebox{0.85}{\includegraphics{#2.eps}}}
  \end{minipage}}}

\title{Classification of 1D and 2D Orbifolds}

\author{Lars Nilse\\
School of Physics and Astronomy, University of Manchester, Manchester M13 9PL, UK\\[10pt]
Email: {\tt lars@hep.man.ac.uk}}

\abstract{We present a complete classification of all 1D and 2D orbifold compactifications. There exist 2 one-dimensional and 17 two-dimensional orbifolds. The classification includes orbifolds such as $S^1/\Zb_2$ or $T^2/\Zb_n$, as well as less familiar ones like $T^2/\Db_n$ or the M\"obius strip. We derive the explicit form of the basis functions and prove their orthonormality and completeness. Our study is based on the classification of space groups, which is well-known from crystallography. We define these groups in a novel, purely algebraic way. That enables us to determine all possible parities that can be defined on the orbifolds. We discuss field theories on $T^2/\Zb_n$ with brane kinetic terms, and describe the derivation of their mass eigenstate bases.}


\preprint{{\tt hep-ph/0601015}}

\begin{document}


\section{Introduction}

In the late 1990s it was realized that extra spatial dimensions are not necessarily linked to the Planck scale and could potentially be relevant for $TeV$ scale physics~\cite{LED}. Since then, extra dimensional models have enjoyed great popularity among physicists. In many of these models, the number of dimensions is reduced by compactifying them on so-called \emph{orbifolds}. Orbifolds are quotient spaces of a manifold modulo a discrete \emph{space group}. In crystallography, the classification of these groups is well-established. There exist two one-dimensional (1D) and 17 two-dimensional (2D) space groups~\cite{CRC}. Based on the classification of these space groups\footnote{The classification has previously been used in a study of conformal field theories~\cite{Dulat:2000xj}.}, we present in this paper a complete classification of 1D and 2D orbifolds. We derive the explicit form of the corresponding basis functions and prove their orthonormality and completeness. Higher-dimensional fields defined on the orbifolds can possess various parities including Scherk-Schwarz (SS) phases. Using a novel \emph{algebraic} definition of the space groups, we derive the set of all possible parities for each of the orbifolds. For example, we show that a complex field compactified on $T^2/\Zb_6$ cannot possess any SS parities. -- In this paper, we present a complete catalogue of all possible 1D and 2D orbifold compactifications. We hope it will prove to be a helpful tool for model builders. The main results are summarized in Tables 1 and 2.

The concept of extra dimensions led to new perspectives in nearly all areas of current \emph{Beyond Standard Model} research~\cite{Csaki:2004ay, reviews}. Take for example \emph{neutrino physics}. 
The observed light neutrino masses are conventionally explained by the seesaw mechanism~\cite{Gell-Mann:1980vs}. It requires a heavy mass scale that might not be available in some of the extra-dimensional scenarios. One possible alternative explanation assumes sterile bulk neutrinos coupling to brane-localized Standard Model (SM) fermions~\cite{neutrinos, Dudas:2005vn}. The light neutrino masses arise now from the volume-suppressed Yukawa couplings. Any neutrino mass model is intimately linked to other research areas, such as leptogenesis or neutrinoless double beta decay~\cite{Bhattacharyya:2002vf}, and therefore very likely to be experimentally testable. Another good example is \emph{electroweak physics}~\cite{electroweak, Gabriel:2004ua}. In the SM, the unitarity of high energy scattering processes is ensured by Higgs exchange. In extra dimensional models, no Higgs is necessary. Unitarity is the consequence of the exchange of an infinite tower of Kaluza-Klein (KK) modes~\cite{Csaki:2003dt, Muck:2004br, LarsPhD}. In a similar way, extra dimensions play an important role in the study of dark matter~\cite{darkmatter} and dark energy~\cite{darkenergy, Ghilencea:2005vm}. -- The majority of the models cited in this paragraph are 5D. They rely on a single extra dimension. Some more recent papers~\cite{Dudas:2005vn, Gabriel:2004ua, Ghilencea:2005vm, further6D} explore the possibilities of 6D models. As we will see, 2D orbifold compactifications are more varied and possess a richer structure. They can also be employed in string compactifications~\cite{string}.

The paper is organized as follows. In Sections~\ref{1D} and~\ref{2D}, we discuss in detail each of the 1D and 2D orbifolds. We derive the basis functions and study the possible parities of complex fields on these orbifolds. In Appendix~\ref{sumrules}, we revisit our study of high energy unitarity on $S^1/\Zb_2$~\cite{Muck:2004br, LarsPhD}. We present a simplified proof of important sum rules that were central in our discussion. The proof relies only on the orthonormality and completeness of the basis and can therefore easily be adapted for any of the orbifolds discussed in this paper. In Appendix~\ref{SSonT2Z2}, we discuss Scherk-Schwarz phases on 2D orbifolds that were previously ignored in Section~\ref{2D}. In Appendix~\ref{BKTonT2Z3}, we describe how to derive the mass eigenstate bases for quantum field theories compactified on $T^2/\Zb_n$ with brane kinetic terms (BKT) at the orbifold fixed points. Some important compactifications such as $S^1/(\Zb_2 \times \Zb_2')$ \cite{Hebecker:2001wq}, $T^2/(\Zb_2 \times \Zb_2' \times \Zb_2'')$ \cite{Asaka:2002nd} or the Chiral Square \cite{Burdman:2005sr, Dobrescu:2004zi} do not appear in Tables~1 and~2. They are being discussed in Section~\ref{S1Z2}, Appendix~\ref{SSonT2Z2} and Section~\ref{T2Z4} respectively.


\newpage
\section{1D Orbifolds}
\label{1D}

One-dimensional orbifolds are very simple. There are only two of them, the circle and the interval. Higher-dimensional fields compactified on them can be expanded in terms of exponentials and sines/cosines respectively. Despite this simplicity we present here a more detailed discussion in order to prepare for the arguments of Section~\ref{2D}.

\emph{Orbifolds} are quotient spaces of a manifold modulo a discrete group\footnote{A formal definition can be found in William Thurston's lecture notes~\cite{ThurstonURL}. The terminology \emph{orbifold} was first used in one of his lecture courses in 1976. The concept itself occurs first in 1956 as \emph{V-manifold} in \cite{satake1, satake2}.}. We are not free to choose any arbitrary discrete group, but are restricted to so-called space groups. An n-dimensional (nD) \emph{space group} is defined as a cocompact discrete group of isometries of $\Rb^n$ \cite{Thurston}. These groups are also known as crystallographic groups or Bieberbach groups, and in 2D as wallpaper groups. Their classification\footnote{For $n=1, \, 2, \, 3, \, 4, \, 5, \, 6$ there are $2, \, 17, \, 230, \, 4 \, 895, \, 222 \, 018, \, 28 \, 927 \, 922$ space groups respectively~\cite{Schattschneider, Brown, Plesken}.} is known for dimensions $n \leq 6$.

In the one-dimensional case, we consider quotient spaces $\Rb/\Gamma$ where $\Gamma$ stands for one of the two 1D space groups $\Zb$ and $\Db_\infty$, see Table 1. The most intuitive way of defining the groups is by identifying the isometries of $\Rb$ that act as generators of the group\footnote{It is not difficult to convince onself that there are no further 1D space groups apart from $\Zb$ and $\Db_\infty$. Given a fundamental 'pattern'~$\rightarrow$, there exist only two possible 1D 'wallpapers', $\rightarrow \, \rightarrow \, \rightarrow \, \rightarrow$ and $\rightarrow \, \leftarrow \, \rightarrow \, \leftarrow$.}, see Fig. 1(a) and 1(b). Alternatively, the groups can be characterized in a purely algebraic way.
\begin{table}[h!]
\begin{center}
\begin{minipage}[t]{0.7\textwidth}
\begin{center}
\vspace{0.5cm}
\begin{tabular}{|c||c|c|c|}
\hline
$\Zb$ & $\Rb/\Zb$ & $S^1$ & circle \\
$\Db_\infty$ & $\Rb/\Db_\infty$ & $S^1/\Zb_2$ & interval\\
\hline
\end{tabular}
\caption{The first column lists the 1D space groups $\Gamma$. The following three columns state the corresponding quotient spaces $\Rb/\Gamma = S^1/\Gamma'$ and their geometry.}
\end{center}
\end{minipage}
\end{center}
\end{table}
\begin{align}
\label{Z}
\Zb &= \langle t \rangle\\
\label{Z2}
\Zb_2 &= \langle r| r^2=\id \rangle\\
\label{Doo}
\Db_\infty &= \langle t,r | r^2=\id, (tr)^2=\id \rangle\\
&\supseteq \Zb, \Zb_2 \nonumber
\end{align}
The set of generators and the relations among them define uniquely the structure of the groups. Note that $\Db_\infty$ is a natural extension of the definition of the dihedral group $\Db_n$ (\ref{Dn}) that we will encounter in Section~\ref{2D}. The advantage of this representation-independent definition will become apparent when we discuss the possible parities on these orbifolds. It is important to note that the choice of generator in the definitions is not unique. For example, the space group $\Db_\infty$ can be defined equally well in terms of two $\pi$-rotations\footnote{The two rotations $r$ and $r'$ do not commute, $[r,r'] \neq 0$. Hence, the \emph{free product} $*$ and not the cross product $\times$ appears in (\ref{DooII}).}, see Fig.~1(c).
\begin{equation}
\begin{split}
\label{DooII}
\Db_\infty &\simeq \Zb_2 * \Zb_2\\
&= \langle r| r^2=\id \rangle * \langle r'| r'^2=\id \rangle \qquad \text{with} \quad r' \equiv tr
\end{split}
\end{equation}
Let $\Gamma' \subseteq \Gamma$ be the largest subgroup of $\Gamma$ that does not include translations. By writing the 1D orbifolds as
\begin{equation}
\label{1Dphilosophy}
\Rb/\Gamma = S^1/\Gamma'
\end{equation}
we make contact with the standard notation. The circle $S^1$ and the interval $S^1/\Zb_2$ are the only one-dimensional orbifolds, since the classification of 1D space groups is complete. The notation in the above paragraphs might appear unfamiliar, but the relations in them are well known. Equations~(\ref{Doo}) and (\ref{DooII}) correspond to the results (3.42) and (3.43) in~\cite{Csaki:2004ay}. Note that the circle in Fig. 1(a) has got the radius
\begin{equation}
\label{R}
R=\frac{1}{2 \pi} \; .
\end{equation}
Here and in the rest of the paper, (\ref{R}) sets our scale. That will allow us to write many expressions in a particular simple form. In the two sections below, we will discuss complex scalar fields $\vp(x,y)$ on 1D orbifolds. We will omit the standard four-dimensional spacetime coordinates $x$ in our notation and specify only the dependence on the extra dimension $y$, i.e. $\vp(y)$.
\begin{figure}[!htbp]
  \begin{center}
  \begin{minipage}[t]{0.85\textwidth}
  \centering
  \vspace{1.2cm}
  \Orbifold{$S^1 = \Rb/\Zb$}{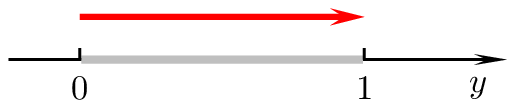}%
  \Orbifold{$S^1/\Zb_2 = \Rb/\Db_\infty$}{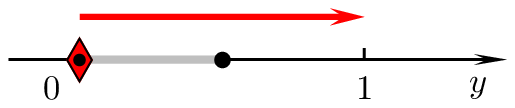}\\[15pt]
  \hspace{0.45\textwidth} \Orbifold{$S^1/\Zb_2 = \Rb/\Db_\infty$}{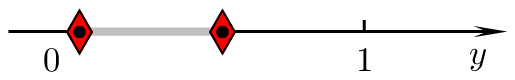}
  \caption{Fundamental domain of the orbifold (thick gray line). Orbifold fixed points (black dots). Generators of the 1D space group: translations (solid red arrows), $\pi$-rotation (red diamond).}
  \label{figS1}
  \end{minipage}
  \end{center}
\end{figure}


\subsection{Circle $S^1$}
\label{S1}

We construct the quotient space $S^1 = \Rb/\Zb$ by identifying the points
\begin{equation}
\label{Z_t}
y \sim y+1
\end{equation}
on $\Rb$, see Fig. 1(a). The shift (\ref{Z_t}) is a representation of the generator $t$ in (\ref{Z}).
We consider complex fields $\vp(y)$ on the circle $S^1$ and allow for Scherk-Schwarz (SS) phases. It is the higher-dimensional Lagrangian $\Lag_{\operatorname{5D}}$ and not the individual fields $\vp(y)$ that we demand to be invariant under transformations (\ref{Z_t}). In the majority of cases, the Lagrangian will include bilinear terms, say kinetic terms $\vp(y) \vp^*(y) \subset \Lag_{\operatorname{5D}}$. We therefore restrict the possible SS phases to
\begin{equation}
p \equiv \exp[i 2 \pi \rho] \qquad \text{with} \quad \rho \in [0,1) \subset \Qb \; .
\end{equation}
A complex field $\vp(y)$ with SS parity $p$ can be expanded in terms of basis functions $f^{(p)}_k(y)$.
\begin{align}
\vp(y+1) &= \exp[i 2 \pi \rho] \, \vp(y)\\
\label{S1expansion}
\vp(y) &= \sum_{k=-\infty}^{\infty} \; \vp_{(k)} f^{(p)}_k(y)\\[10pt]
\label{fk}
f^{(p)}_k(y) &= \exp[i2 \pi (k + \rho) y]\\
&= \exp[i M^{(p)}_k y] \qquad \text{with} \quad M^{(p)}_k \equiv 2 \pi (k+\rho)
\end{align}
In the following, let $j \in \Nb$ be the smallest integer\footnote{In the case that several complex fields with distinct parities $p_n \equiv \exp[i 2 \pi \rho_n]$ appear in the higher-dimensional Lagrangian $\Lag_{\operatorname{5D}}$, let $j$ be the the smallest integer such that $p^j_n=1$ for all $n$. The rational $\rho_n \subset \Qb$ ensure that there always exists such a $j$.} such that $p^j=1$. The set of basis functions is orthonormal in the following sense.
\begin{equation}
\label{S1orth}
\int_0^j \frac{dy}{j} \; f^{(p)}_k(y) f^{(q) *}_l(y) = \delta_{k,l} \delta_{p,q}
\end{equation}
We have seen that the translation (\ref{Z_t}) sets our scale. It is therefore natural to find the associated parity $p$ to appear in the spectrum $m_k^{(p)}$. In the expressions below, $\partial_5 \equiv \partial/\partial y$ stands for the derivative with respect to the compactified dimension.
\begin{align}
\label{S1trans}
f^{(p)}_k(y+1) &= \exp[i 2 \pi \rho] \, f^{(p)}_k(y)\\
\label{S1deriv}
\partial_5 f^{(p)}_k(y) &= i M^{(p)}_k f^{(p)}_k(y)\\
\label{S1wave}
\bigl[ \partial_5^2 + m_k^{(p)2}\bigr] f^{(p)}_k(y) &= 0\\
\text{with} \quad m_k^{(p)2} &\equiv M^{(p)}_k M^{(p)*}_k = \bigl[2 \pi \; (k+\rho) \bigr]^2 \nonumber
\end{align}
There exists an interesting relation between argument and indices of the exponentials. It will prove to be central for the derivation of the $S^1/\Zb_2$ basis function in the next section.
\begin{equation}
\label{S1sym}
f^{(p)}_k(-y) = f^{(p^*)}_{-k}(y)
\end{equation}


\subsection{Interval $S^1/\Zb_2$}
\label{S1Z2}

The interval $S^1/\Zb_2$ is the quotient space of the real line modulo $\Db_\infty$. The two generators of the discrete group $\Db_\infty$ (\ref{Doo}) are represented by the transformations
\begin{align}
\label{Doo_t}
y &\sim y+1\\
\label{Doo_r}
&\sim -y
\end{align}
in $\Rb$. The compactification radius (\ref{R}) sets again our scale. The orbifold possesses two distinct fixed points at $y=0$ and $y=1/2$. They are invariant under group elements $r$ and $tr=r'$ respectively. Unlike the circle, the orbifold $S^1/\Zb_2$ puts restrictions on the possible parities that can be defined on it. Any assignment of parities has to be consistent with the defining relations of $\Db_\infty$. Solving (\ref{Doo}) in $\Cb$, we find four solutions.
\begin{equation}
\label{Doo_C}
t=\pm \; , \quad r=\pm
\end{equation}
Whereas (\ref{Doo_t}) and (\ref{Doo_r}) are a representation of $\Db_\infty$ in the space of isometric maps in $\Rb$, (\ref{Doo_C}) is a representation\footnote{The same symbols $t$ and $r$ are used for the group elements in (\ref{Doo}) and their specific representation in $\Cb$.} in $\Cb$. A complex field $\vp(y)$ on the interval can therefore posses four different parities $(p,q)$ with $p,q=\pm$.
\begin{align}
\label{S1Z2_r}
\vp(-y) &= p \, \vp(y)\\
\label{S1Z2_t}
\vp(y+1) &= q \, \vp(y)
\end{align}
Note that we make a clear distinction between the definition of the orbifold, (\ref{Doo_t}) and (\ref{Doo_r}), and any physics (\ref{S1Z2_r}) and (\ref{S1Z2_t}) defined on it. The field $\vp(y)$ can be expanded in terms of basis functions $F^{(p,q)}_k(y)$. Using (\ref{fk}) and (\ref{S1sym}), we find their form.
\begin{align}
\label{S1Z2_F}
F^{(p,q)}_k(y) &= c_k^{(p,q)} \bigl[ f^{(q)}_k(y) + p f^{(q)}_{-k}(y) \bigr]\\
\label{S1Z2_FII}
&=\begin{cases}
\sqrt{2^{1-\delta_{k,0}}} \; \cos (2 k \pi y) \qquad &\text{for} \quad (p,q)=(+,+)\\
\sqrt{2^{1-\delta_{k,0}}} \; \cos ([2k+1] \pi y) \qquad &\text{for} \quad (p,q)=(+,-)\\
\sqrt{2} i \; \sin (2 k \pi y) \qquad &\text{for} \quad (p,q)=(-,+)\\
\sqrt{2^{1-\delta_{k,0}}} i \; \sin ([2k+1] \pi y) \qquad &\text{for} \quad (p,q)=(-,-)
\end{cases}
\end{align}
Since there are only two possible SS phases $t=\pm$, we use $j=2$ in (\ref{S1orth}). The basis is orthonormal in the following sense.
\begin{equation}
\label{S1Z2orth}
\int_0^{2} \frac{dy}{2} \; F^{(p,q)}_k(y) F^{(r,s)*}_l(y) = \delta_{k,l} \delta_{p,r} \delta_{q,s}
\end{equation}
On $S^1/\Zb_2$, there exist four parities. In (\ref{S1Z2orth}) we integrate four times over the fundamental domain of the orbifold $[0,1/2]$, i.e. $[0,2]$. On $S^1$, we can work with $j$ different parities $\{+, p, p^2, \ldots, p^{j-1}\}$ and integrate $j$ times over the fundamental domain $[0,1)$, i.e. $[0,j)$. In general, we always integrate over a region in $\Rb$ over which any of the basis functions is guaranteed to be periodic. Using (\ref{S1orth}), we can check the above relation and determine the normalization constants.
\begin{equation}
c^{(p,q)}_k = \sqrt{2^{-1-\delta_{k,0}}}
\end{equation}
Using (\ref{S1sym}) and one of the orbifold symmetries in (\ref{S1Z2_F}), we find
\begin{equation}
F^{(\pm,q)}_k(-y)=F^{(\pm,q)}_{-k}(y)=\pm F^{(\pm,q)}_k(y) \; .
\end{equation}
The second of the above equalities implies that the basis functions are not independent. Unlike in (\ref{S1expansion}), we have to restrict the coefficient in the expansion of the complex field, $k\geq0$.
\begin{equation}
\label{S1Z2expansion}
\vp(y) = \sum_{k=0}^{\infty} \; \vp_{(k)} F^{(p,q)}_k(y)
\end{equation}
Using (\ref{S1deriv}) and (\ref{S1wave}), we derive the relations below and find the spectrum $m_k^{(q)}=M^{(q)}_k$. The distinction between $m_k^{(q)}$ and $M^{(q)}_k$ will become relevant in Section~\ref{2D}, where $M^{(q)}_k$ can be complex.
\begin{align}
F^{(\pm,q)}_k(-y) &= \pm F^{(\pm,q)}_k(y)\\
F^{(p,\pm)}_k(y+1) &= \pm F^{(p,\pm)}_k(y)\\
\label{S1Z2deriv}
\partial_5 F^{(\pm,q)}_k(y) &= i M^{(q)}_k F^{(\mp,q)}_k(y)\\
\label{S1Z2wave}
\bigl[ \partial_5^2 + m_k^{(q)2} \bigr] F^{(p,q)}_k(y) &=0 \qquad \text{with} \quad m^{(q)2}_k\equiv M^{(q)}_k M^{(q)*}_k\\
M^{(q)}_k &\equiv \begin{cases} 2 \pi \, k \qquad &\text{for} \quad q=+\\ 2 \pi \, (k+1/2) \qquad &\text{for} \quad q=-\end{cases}
\end{align}
Following our discussion in~\cite{Muck:2004br, LarsPhD}, we construct the expansion of the delta function on $S^1/\Zb_2$. Substituting an even ansatz in the defining relation (\ref{DefDelta}), we find (\ref{S1Z2delta}).
\begin{equation}
\label{DefDelta}
\int_0^2 \frac{dy}{2} \; \vp(y) \delta^*(y-y') = \vp(y')
\end{equation}
\begin{align}
\delta(y) &= \sum_{k=0}^\infty F^{(+,+)*}_k(0) F^{(+,+)}_k(y)\\
\label{S1Z2delta}
&= \sum_{k=0}^{\infty} \; 2^{1-\delta_{k,0}} \, \cos (2 \pi k y)
\end{align}
Using the explicit expressions (\ref{S1Z2_FII}) and (\ref{S1Z2delta}), we can check the completeness of our basis. Alternatively, we can verify the consistency of orthonormality, completeness and delta function by substituting (\ref{S1Z2complete}) in (\ref{DefDelta}).
\begin{equation}
\label{S1Z2complete}
\delta (y_1-y_2) = \sum_{p,q=\pm}\sum_{n=0}^{\infty} \, F^{(p,q)}_n (y_1) \, F^{(p,q)*}_n (y_2)
\end{equation}
It is often convenient to work with a real basis. We can drop the $i$ in (\ref{S1Z2_FII}) and hence the complex conjugation in the orthonormality (\ref{S1Z2orth})\footnote{Here we prefer (\ref{S1Z2_FII}), since (\ref{S1Z2deriv}) is of the same form for all parities.}. In that case, the basis becomes identical to (4)-(7) in \cite{Hebecker:2001wq}\footnote{The basis in \cite{Hebecker:2001wq} is normalized but not entirely orthogonal in $[0,\pi R /2]$. Deviding (4)-(7) by a factor 2, the basis becomes completely orthonormal in $[0,2 \pi R]$. In the final paragraph of Appendix~\ref{sumrules}, we explain why we find it necessary to work with the slightly modified orthonormality~(\ref{S1Z2orth}) instead of the one used in~\cite{Hebecker:2001wq}. -- Transferring the normalization factor $1/2$ of our integration measure to the basis functions, we divide (\ref{S1Z2_FII}) by $\sqrt{2}$.} with $R=1/\pi$. Fields of parity $(p,q)$ on an orbifold $S^1/\Zb_2$ with radius $R$ correspond to parities $(p,pq)$ on $S^1/(\Zb_2 \times \Zb_2')$ with radius $2R$. Earlier on we discussed the fact that the choice of the generators of $\Db_\infty$ is not unique. In (\ref{S1Z2_r}) and (\ref{S1Z2_t}), we assign parities to $r$ and $t$, whereas the authors of \cite{Hebecker:2001wq} work with $r$ and $r'=tr$ instead, see Fig. 1(b) and 1(c). We suggest that $S^1/(\Zb_2 \times \Zb_2')$ is not a separate orbifold, but indeed $\Rb/(\Zb_2 * \Zb_2)=S^1/\Zb_2$ with non-trivial SS phases.

Let us consider the case of trivial SS phases, i.e. our theory includes only fields with parities $(\pm,+)$. The orthonormality (\ref{S1Z2orth}) can then be simplified by changing the upper boundary to $1$ and removing the normalization factor $1/2$ in the integration measure. Our basis becomes then identical to (B.15) in \cite{LarsPhD} with radius $R=1/(2 \pi)$. 


\section{2D Orbifolds}
\label{2D}

In the previous section, we merely reviewed the known 1D orbifolds. In our discussion below, we will describe a number of two-dimensional orbifolds that have rarely been used in the literature. We hope that model builders will find them interesting and useful.

We consider 2D orbifolds as quotient spaces $\Rb^2/\Gamma$ where $\Gamma$ stands for one of the 17 two-dimensional space groups listed in Table 2. In this paper, we adopt the crystallographic notation~\cite{CRC, Schattschneider} for the space groups. Conway's orbifold notation~\cite{Conway} is more common among mathematicians and is listed in the second column of Table 2. The 2D space groups are defined in Figures 2 to 5 by specifying the isometries of $\Rb^2$ that act as generators. The structures in the plane are much richer than on $\Rb$. The possible isometries include translations, reflections, $2 \pi / n$-rotations with $n=2,3,4,6$ and so-called glide-reflections, which are translations with a simultaneous mirror reflection. By acting with the generators on the fundamental domain of the orbifold, it is possible to map the entire plane. Note that the choice of generators in the definitions is not unique, cf. Chart 5 in~\cite{Schattschneider}.

As done in Section~\ref{1D}, we can also define the space groups in a purely algebraic way. Instead of specifying a particular representation of the generators, we list the relations among them and thereby describe the structure of the groups.
\begin{table}[h!]
\begin{center}
\begin{minipage}[t]{0.9\textwidth}
\begin{center}
\begin{tabular}{|c|c||c|c|c|c|}
\hline
$\pI$ & $\circ$ & $\Rb^2/\Zb^2$ & $T^2$ & torus & $r, \theta$\\
$\pg$ & $\times \times$ & $\Rb^2/\pg$ & & Klein bottle & $r$\\
$\pgg$ & $22 \times$ & $\Rb^2/\pgg$ & $\Rb P^2$ & real projective plane & $r$\\
\hline
$\pII$ & $2222$ & $\Rb^2/\pII$ &$T^2/\Zb_2$ & 4-pillow & $r, \theta$\\
$\pM$ & $**$ & $\Rb^2/(\Zb \times \Db_\infty)$ &$T^2/\Zb'_2$ & annulus & $r$\\
$\cm$ & $* \times $ & $\Rb^2/\cm$ &$T^2/\Zb''_2$ & M\"obius strip & $\theta$\\
$\pmm$ & $*2222$ & $\Rb^2/\Db_\infty^2$ &$T^2/\Db_2$ & rectangle & $r$\\
$\cmm$ & $2*22$ & $\Rb^2/\cmm$ &$T^2/\Db'_2$ & triangle & $\theta$\\
$\pmg$ & $22*$ & $\Rb^2/\pmg$ &$T^2/ \Fb_2$ & open 4-pillow & $r$\\
\hline
$\pIII$ & $333$ & $\Rb^2/\pIII$ & $T^2/\Zb_3$ & 3-pillow & \\
$\pIIImI$ & $*333$ & $\Rb^2/\pIIImI$ & $T^2/ \Db_3$ & triangle & \\
$\pIIIIm$ & $3*3$ & $\Rb^2/\pIIIIm$ & $T^2/ \Fb_3$ & open 3-pillow & \\
\hline
$\pIV$ & $442$ & $\Rb^2/\pIV$ & $T^2/\Zb_4$ & 3-pillow & \\
$\pIVm$ & $*442$ & $\Rb^2/\pIVm$ & $T^2/ \Db_4$ & triangle & \\
$\pIVg$ & $4*2$ & $\Rb^2/\pIVg$ & & open 3-pillow & \\
\hline
$\pVI$ & $632$ & $\Rb^2/\pVI$ & $T^2/\Zb_6$ & 3-pillow & \\
$\pVIm$ & $*632$ & $\Rb^2/\pVIm$ & $T^2/ \Db_6$ & triangle & \\
\hline
\end{tabular}
\caption{The first two columns list the 2D space groups $\Gamma$ in crystallographic~\cite{CRC} and orbifold~\cite{Conway} notation. The following three columns state the corresponding quotient spaces $\Rb^2/\Gamma = T^2/\Gamma'$ and their geometry. In the last column we lists any free parameters that specify the shape of the space.}
\end{center}
\end{minipage}
\end{center}
\end{table}
\begin{align}
\label{p1}
\pI &\simeq \Zb^2\\
&= \langle t_1 \rangle \times \langle t_2 \rangle \nonumber\\
\label{pg}
\pg &= \langle t_2, g | [g^2,t_2]=0, t_2 g t_2 g^{-1} = \id \rangle\\
&\supseteq \Zb^2 \nonumber\\
\label{pgg}
\pgg &= \langle r,g | r^2=(g^2r)^2=\id \rangle\\
&\supseteq \Zb^2, \Zb_2 \nonumber\\[6mm]
\label{p2}
\pII &= \langle t_1, t_2, r | r^2= (t_1 r)^2 = (r t_2)^2 =\id, [t_1,t_2]=0 \rangle\\
&\supseteq \Zb^2, \Db_\infty, \Zb_2 \nonumber\\
\label{pm}
\pM &\simeq \Zb \times \Db_\infty\\
&= \langle t_1 \rangle \times \langle t_2, f| f^2=(t_2 f)^2=\id \rangle \nonumber\\
&\supseteq \Zb^2, \Db_\infty, \Zb_2 \nonumber\\
\label{cm}
\cm &= \langle t_1, t_2, f | f^2=\id, [t_1,t_2]=0, f t_1 = t_2 f \rangle\\
&\supseteq \Zb^2, \Zb_2 \nonumber
\end{align}
\begin{align}
\label{pmm}
\pmm &\simeq \Db_\infty^2\\
&= \langle t_2, f| f^2=(t_2 f)^2=\id \rangle \times \langle t_1 t_2, r | r^2=(t_1 t_2 r)^2=\id \rangle \nonumber\\
&\supseteq  \Zb^2, \pII, \Db_\infty, \Db_2, \Zb_2 \nonumber\\
\label{cmm}
\cmm &=\langle t_1, t_2, r, f | r^2=f^2=(f r)^2=(t_1 r)^2 = (r t_2)^2 =\id, [t_1,t_2]=0, f t_1 = t_2 f \rangle\\
&\supseteq \Zb^2, \pII, \cm, \Db_2, \Zb_2 \nonumber\\
\label{pmg}
\pmg &=\langle t_1, t_2, r, f | r^2=f^2=(t_1 r)^2 = (r t_2)^2 =\id, [t_1,t_2]=0 \rangle\\
&\supseteq \Zb^2, \pII, \Fb_2, \Zb_2 \nonumber\\[5mm]
\label{p3}
\pIII &= \langle t_1, t_2, r | r^3=(t_1 r)^3 = (t_2 r^2)^3 =\id, r t_1 = t_2 r, [t_1,t_2]=0 \rangle\\
&\supseteq \Zb^2, \Zb_3 \nonumber\\
\begin{split}
\label{p3m1}
\pIIImI &= \langle t_1, t_2, r, f | r^3=f^2=(fr)^2=(t_1 r)^3 = (t_2 r^2)^3 =\id,\\
&\qquad r t_1 = t_2 r, f t_1 = t_2 f, [t_1,t_2]=0 \rangle
\end{split}\\
&\supseteq \Zb^2, \pIII, \Db_3, \Zb_3, \Zb_2, \cm \nonumber\\
\begin{split}
\label{p31m}
\pIIIIm &= \langle t_1, t_2, r, f | r^3=f^2=(frfr^2)^3=(t_1 r)^3 = (t_2 r^2)^3 =\id,\\
&\qquad r t_1 = t_2 r, t_1^{-1} f t_1 = t_2^{-1} f t_2, [t_1,t_2]=0 \rangle
\end{split}\\
&\supseteq \Zb^2, \pIII, \Fb_3, \Zb_3, \Zb_2 \nonumber\\[5mm]
\label{p4}
\pIV &= \langle t_1, t_2, r | r^4=(t_1 r)^4=(t_2 r^3)^4=\id, r t_1 = t_2 r, [t_1,t_2]=0 \rangle\\
&\supseteq \Zb^2, \Zb_4, \Zb_2 \nonumber\\
\begin{split}
\label{p4m}
\pIVm &= \langle t_1, t_2, r, f | r^4=f^2=(fr)^2=(t_1 r)^4=(t_2 r^3)^4=\id,\\
&\qquad  r t_1 = t_2 r, f t_1 = t_2 f, [t_1,t_2]=0 \rangle
\end{split}\\
&\supseteq \Zb^2, \pIV, \Db_4, \Zb_4, \Zb_2, \pII, \cm, \cmm \nonumber\\
\label{p4g}
\pIVg &= \langle r,f | r^4=f^2=(frfr^3)^2=\id \rangle\\
&\supseteq \Zb^2, \pIV, \Db_2, \Zb_4, \Zb_2 \nonumber\\[5mm]
\label{p6}
\pVI &= \langle t_1, t_2, r | r^6=\id, \, r t_1 = t_2 r, \, t_1 r^2 t_1 = r t_1 r, \, [t_1,t_2]=0 \rangle\\
&\supseteq \Zb^2, \pIII, \Zb_6, \Zb_3, \Zb_2 \nonumber\\
\begin{split}
\label{p6m}
\pVIm &= \langle t_1, t_2, r, f | r^6=f^2=(fr)^2=\id, \, r t_1 = t_2 r,\\
&\qquad t_1 r^2 t_1 = r t_1 r, \, f t_1 = t_2 f, \, [t_1,t_2]=0 \rangle
\end{split}\\
&\supseteq \Zb^2, \Db_6, \Db_2, \Zb_6, \Zb_3, \Zb_2, \pII, \cm, \cmm \nonumber
\end{align}
In expressions~(\ref{p1}) to (\ref{p6m}) we list all possible subgroups of the space groups. For each $\Gamma$, let $\Gamma' \subseteq \Gamma$ be the largest of the subgroups that does not include translations. The group $\Gamma'$ is either a cyclic group $\Zb_n$, a dihedral group $\Db_n$  or one of the two groups\footnote{There exists no established mathematical notation for these two groups. $\Fb_2$ and $\Fb_3$ are the only new notation that we introduce in this paper.} $\Fb_2$ and $\Fb_3$.
\newpage
\begin{align}
\Zb_n &= \langle r | r^n=\id \rangle\\
\label{Dn}
\Db_n &= \langle r,f | r^n=f^2=(fr)^2=\id \rangle\\
\label{F2}
\Fb_2 &= \langle r,f | r^2=f^2=\id \rangle\\
\label{F3}
\Fb_3 &= \langle r,f | r^3=f^2=(f r f r^2)^3 =\id \rangle
\end{align}
We are now able to rewrite the 2D orbifolds as
\begin{equation}
\label{2Dphilosophy}
\Rb^2/\Gamma = T^2/\Gamma' \; .
\end{equation}
The exceptions are the Klein bottle, the real projective plane $\Rb P^2$ and $\Rb^2/\pIVg$. The translations in the space groups $\pg$, $\pgg$ and $\pIVg$ are not generators but are themselves generated, and hence no non-trivial subgroups $\Gamma'$ exist. We use the $T^2/\Gamma'$ notation to label the two-dimensional orbifolds. It conforms with the standard notation for the orbifolds $T^2/\Zb_n$. -- On the other hand, it is important to stress that the discussion in this paper is based on the $\Rb^2/\Gamma$ picture of 2D orbifolds. The translations that appear as generators in (\ref{p1}) to (\ref{p6m}) differ in no way from the other generators. The SS phases associated with them should not be considered as special, neither should the fundamental domain of the torus.

What do these orbifolds look like? Let us consider the simple example of $\Rb^2/\Zb^2$. Acting with the generators of the space group $\pI$ on the fundamental domain in Fig.~2(a), we realize that opposite edges of the parallelogram are identical. The parallelogram is the surface of a torus $T^2=\Rb^2/\Zb^2$. -- The orbifold $\Rb^2/\cm$ turns out to be a more interesting example. Acting with $t_1 f \in \cm$ on the grey triangle in Fig.~3(e), we identify the two short sides as indicated by the arrows. The third side is invariant under $f \in \cm$ and hence an \emph{orbifold fixed line}\footnote{Compare with the orbifold fixed points $y=0$ and $y=1/2$ in Fig.~1(b), which are invariant under $r$ and $t r \in \Db_\infty$ respectively.}. The triangle is the surface of a M\"obius strip. -- In the same way, we can identify $T^2/\Zb_3$ in Fig.~4(a) with a triangular pillow, or $T^2/\Fb_2$ in Fig.~3(b) with a rectangular pillow with one side slit open. The results of these constructions are summarized in the fifth column of Table~2.

In each of the Figures~2(a) to 5(b) we highlight the fundamental domain of a torus. Let $R_1$ and $R_2$ its compactification radii. The radius $R_1=R=1/(2 \pi)$ which is associated with a translation along the $x_5$ axis sets our scale, see (\ref{R}). The radius $R_2=r R_1$ belongs to a second translation at an angle~$\theta$. The parameters $r$ and $\theta$ determine the shape of the orbifold. The last column of Table~2 list which of them can be chosen freely. -- In what follows, we discuss complex scalar fields $\vp(x,y_5,y_6)$ on 2D orbifolds. In our notation we will omit the four-dimensional spacetime coordinates, i.e. $\vp(y_5,y_6)$.
\vspace{2cm}
\begin{figure}[!htbp]
  \begin{center}
  \begin{minipage}[t]{0.75\textwidth}
  \centering
  \Orbifold{$T^2 = \Rb^2/\pI = \Rb^2/\Zb^2$}{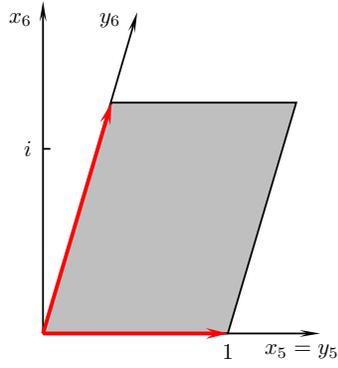} \hspace{0.425\textwidth} \phantom{} \vspace{15mm}
  \Orbifold{$\Rb^2/\pg$}{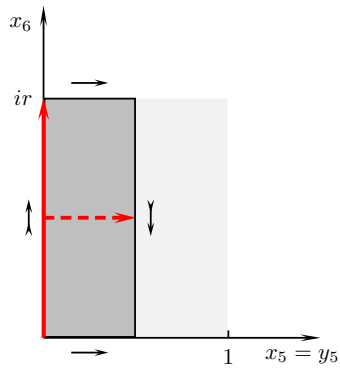}%
  \Orbifold{$\Rb P^2 = \Rb^2/\pgg$}{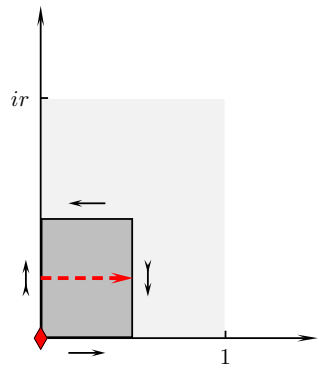}
  \caption{Fundamental domains of the orbifold (gray) and the torus (light gray). Generators of the 2D space group: translations (solid red arrows), glide-reflections (dashed red arrows), $\pi$-rotations (red diamond). In some of the figures, small black arrows indicate which edges of the orbifold fundamental domain have to be identified.}
  \label{figT2}
  \end{minipage}
  \end{center}
\end{figure}
\begin{figure}[!htbp]
  \begin{center}
  \begin{minipage}[t]{0.85\textwidth}
  \centering
  \Orbifold{$T^2/\Zb_2 = \Rb^2/\pII$}{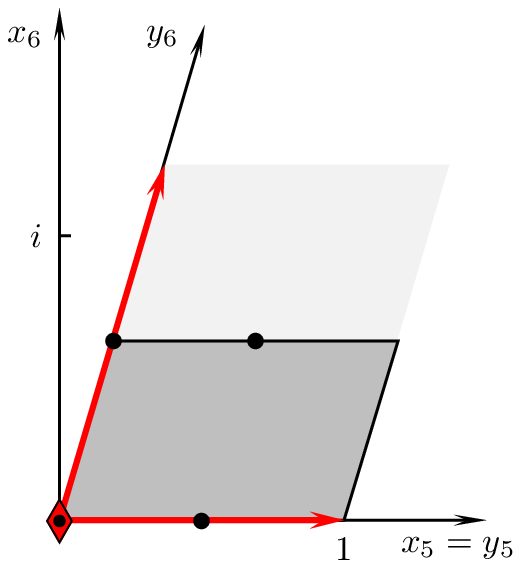}%
  \Orbifold{$T^2/\Fb_2 = \Rb^2/\pmg$}{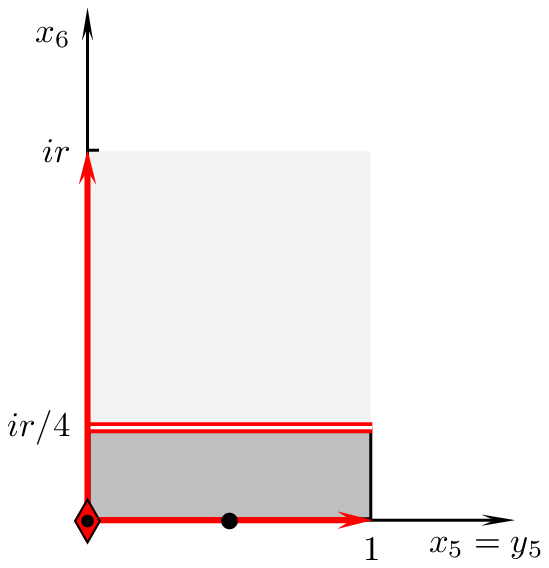}
  \Orbifold{$T^2/\Zb_2' = \Rb^2/\pM = \Rb^2/(\Zb \times \Db_\infty)$}{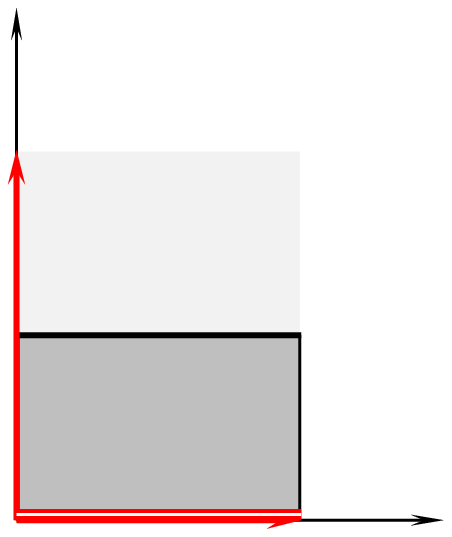}%
  \Orbifold{$T^2/\Db_2 = \Rb^2/\pmm = \Rb^2/\Db_\infty^2$}{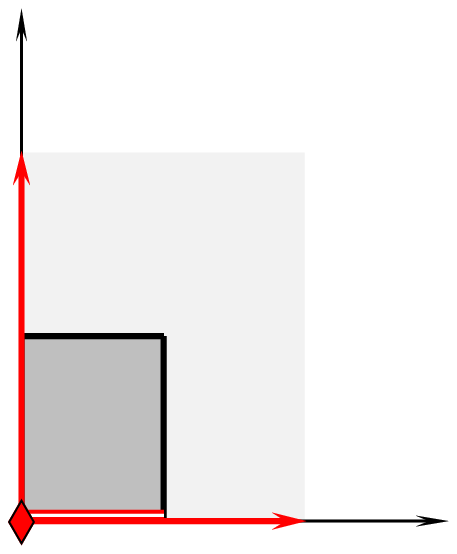} \vspace{6mm}
  \Orbifold{$T^2/\Zb_2'' = \Rb^2/\cm$}{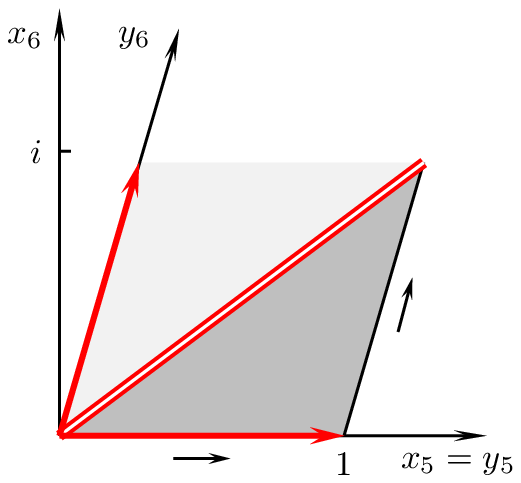}%
  \Orbifold{$T^2/\Db_2' = \Rb^2/\cmm$}{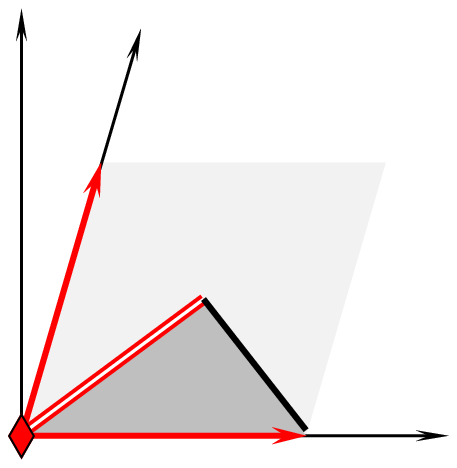}
  \caption{Fundamental domains of the orbifold (gray) and the torus (light gray). Generators of the 2D space group: translations (solid red arrows), reflections (double red lines), $\pi$-rotations (red diamond). Fixed points (black dots), fixed lines (strong black lines or double red lines).}
  \label{figT2Z2}
  \end{minipage}
  \end{center}
\end{figure}
\begin{figure}
  \begin{center}
  \begin{minipage}[t]{0.95\textwidth}
  \centering
  \Orbifold{$T^2/\Zb_3 = \Rb^2/\pIII$}{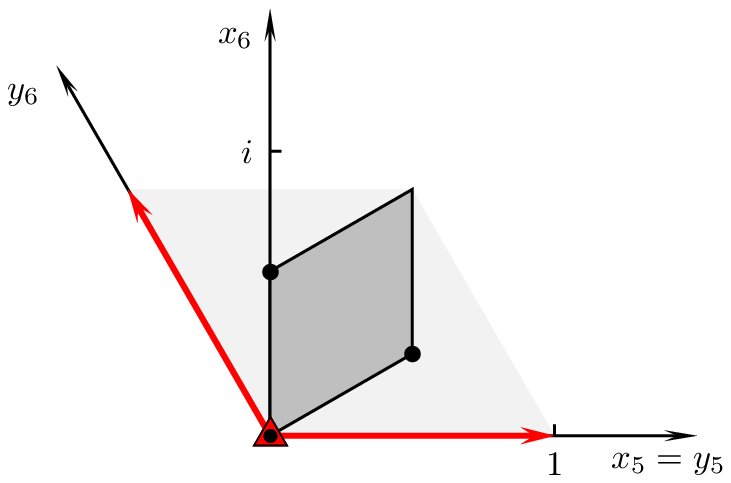}%
  \Orbifold{$T^2/\Zb_4 = \Rb^2/\pIV$}{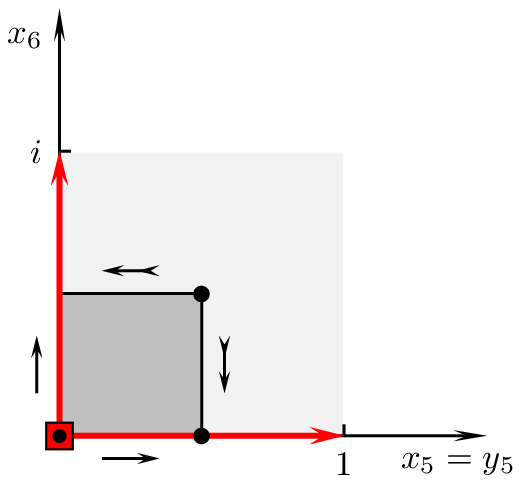} \vspace{6mm}
  \Orbifold{$T^2/\Db_3 = \Rb^2/\pIIImI$}{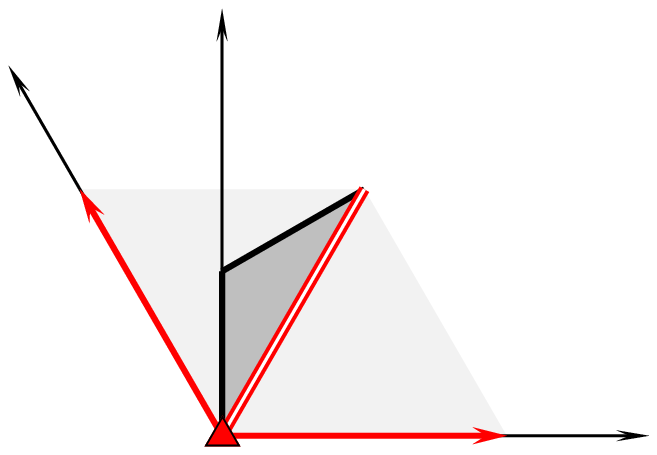}%
  \Orbifold{$T^2/\Db_4 = \Rb^2/\pIVm$}{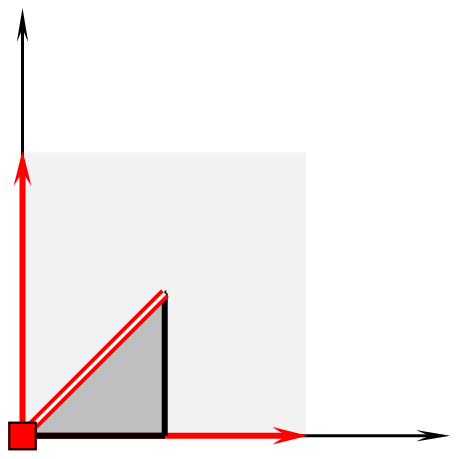} \vspace{6mm}
  \Orbifold{$T^2/\Fb_3 = \Rb^2/\pIIIIm$}{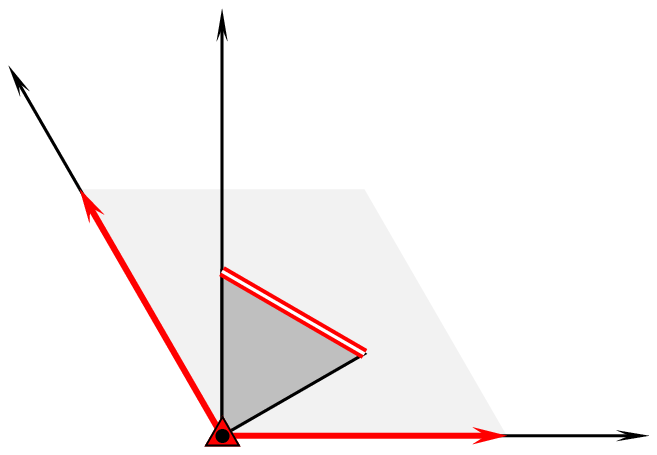}%
  \Orbifold{$\Rb^2/\pIVg$}{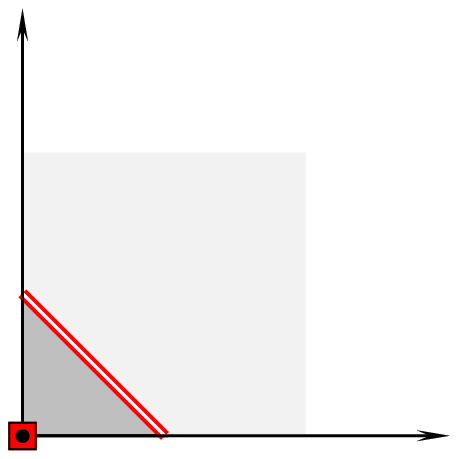}
  \caption{Fundamental domains of the orbifold (gray) and the torus (light gray). Generators of the 2D space group: translations (solid red arrows), reflections (double red lines), $2 \pi/3$-rotations (triangle), $\pi /2$-rotations (square). Fixed points (black dots), fixed lines (strong black lines or double red lines).}
  \label{figT2Z3}
  \end{minipage}
  \end{center}
\end{figure}
\begin{figure}[!htbp]
  \begin{center}
  \begin{minipage}[t]{0.75\textwidth}
  \centering
  \Orbifold{$T^2/\Zb_6 = \Rb^2/\pVI$}{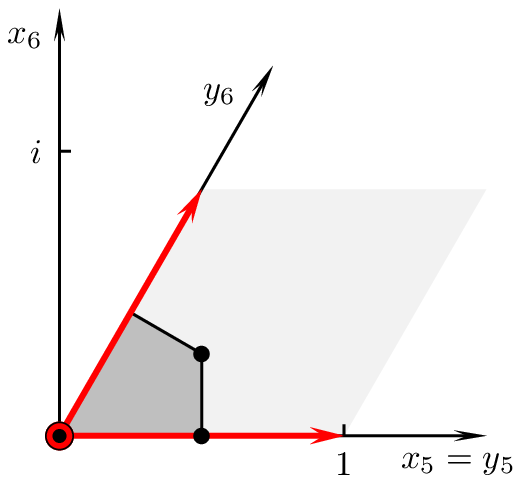}%
  \Orbifold{$T^2/\Db_6 = \Rb^2/\pVIm$}{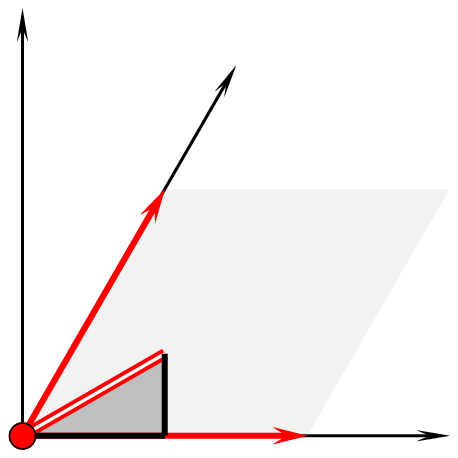}
  \caption{Fundamental domains of the orbifold (gray) and the torus (light gray). Generators of the 2D space group: translations (solid red arrows), reflections (double red lines), $\pi/3$-rotations (red circle). Fixed points (black dots), fixed lines (strong black lines or double red lines).}
  \label{figT2Z6}
  \end{minipage}
  \end{center}
\end{figure}


\newpage
\phantom{}
\newpage
\phantom{}
\newpage
\subsection{Torus $T^2$}
\label{T2}

The space group $\pI \simeq \Zb^2$ (\ref{p1}) is generated by two translations $t_1$ and $t_2$. In the complex plane they correspond to the transformations
\begin{align}
\label{p1_t1}
z &\sim z+1\\
\label{p1_t2}
&\sim z+\omega \; ,
\end{align}
where $R_1=R=1/(2 \pi)$ and $R_2=r R_1$ are the two compactification radii and $\omega=r e^{i \theta}$. The form of any SS phases is merely restricted by the invariance of bilinear terms in the higher-dimensional Lagrangian $\Lag_{\operatorname{6D}}$ under transformations (\ref{p1_t1}) and (\ref{p1_t2})\footnote{The symbols $t_1$ and $t_2$ can stand either for the group elements in (\ref{p1}) or their representation in $\Cb$.}.
\begin{equation}
t_1 = \exp[i 2 \pi \rho_1], \quad t_2 = \exp[i 2 \pi \rho_2] \qquad \text{with} \quad \rho_{1,2} \in [0,1) \subset \Qb
\end{equation}
We delay a detailed discussion of SS phases on 2D orbifolds to Appendix~\ref{SSonT2Z2}. In the following sections, we ignore them $t_1=t_2=+$ and restrict our study to periodic complex fields.
\begin{align}
\vp(z+1) &= \vp(z)\\
\vp(z+\omega) &= \vp(z)
\end{align}
They can be expanded in terms of exponentials $f_{k,l}(z)$\footnote{In what follows, we write any function $g(z,z^*)$ in the complex plane short as $g(z)$. Our notation does \emph{not} imply these functions to be analytic.}. Let $(x_5,x_6)$ be standard Cartesian coordinates and $(y_5,y_6)$ coordinates along the translations $t_1$ and $t_2$, see Fig.~2(a).
\begin{align}
z &= x_5 + i x_6\\
&= y_5 + \omega y_6
\end{align}
\begin{align}
y_5 &= \frac{z+z^*}{2} - \frac{\omega+\omega^*}{\omega-\omega^*} \; \frac{z-z^*}{2}\\
y_6 &= \frac{z-z^*}{\omega-\omega^*}
\end{align}
\begin{align}
f_{k,l}(z) &= \exp \bigl[ i 2 \pi (k y_5 + l y_6) \bigr]\\
\label{fkl}
&=\exp \bigl[i 2 \operatorname{Re}(M_{k,l} z)\bigr]\\
\label{M}
M_{k,l} &= 2 \pi \; \frac{l-\omega^* k}{\omega-\omega^*}
\end{align}
\begin{equation}
\vp(z) = \sum_{k,l=-\infty}^{\infty} \; \vp_{(k,l)} f_{k,l}(z)
\end{equation}
The orthonormality of the basis functions $f_{k,l}(z)$ can be checked in $(y_5,y_6)$ coordinates. In the subsequent sections, we will prefer the complex notation (\ref{T2orth}). The world volume of the compact dimensions is given by $r \sin \theta = (\omega^*-\omega) \; i/2$. For the integration measure we find $dy_5 \, dy_6 = dz \, dz^* / (\omega^*-\omega)$.
\begin{gather}
\int_0^1 \int_0^1 \, \frac{dy_5 dy_6}{r \, \sin \theta} \; f_{k,l}(y_5,y_6) f^*_{m,n}(y_5,y_6) = \delta_{k,l} \delta_{m,n}\\
\label{T2orth}
\int \dzz \; f_{k,l}(z) f^*_{m,n}(z) = \delta_{k,l} \delta_{m,n}
\end{gather}
The basis functions are periodic and obey the following relations.
\begin{gather}
f_{k,l}(z+1) = f_{k,l}(z+\omega) = f_{k,l}(z)\\
\label{T2deriv}
\begin{split}
\pz f_{k,l}(z) &= i M_{k,l} \, f_{k,l}(z)\\
\pzc f_{k,l}(z) &= i M^*_{k,l} \, f_{k,l}(z)
\end{split}
\end{gather}
We recover the spectrum derived in~\cite{Dienes:2001wu}. Apart from $T^2$, it is also characteristic for $T^2/\Zb_2$.
\begin{align}
\label{T2wave}
\bigl[ \pz \pzc + m^2_{k,l}\bigr] f_{k,l}(z) &= 0\\
\label{m2}
m^2_{k,l} &\equiv M_{k,l} M^*_{k,l}\\
&= (2 \pi)^2 \; \frac{l^2 + \omega \omega^* k^2 - (\omega + \omega^*) k l}{2 \omega \omega^* - \omega^2 -\omega^{*2}}\\
&= \frac{1}{\sin^2 \theta} \, \Bigl[ \frac{k^2}{4 R_1^2}+\frac{l^2}{4 R_2^2} - \frac{\cos \theta}{2 R_1 R_2} k l \Bigr]
\end{align}
As we will see in the following sections, the spectra on different orbifolds correspond to specific choices for the complex modulus $\omega$.
\begin{itemize}
\item{$\omega=r i$}\\
The spectrum is characteristic for $T^2/\Zb'_2$, $T^2/\Db_2$, $T^2/\Fb_2$, the Klein bottle and the real projective plane $\Rb P^2$.
\begin{align}
\label{Mri}
M_{k,l} &= \pi [ k -i l/r]\\
\label{m2ri}
m^2_{k,l} &= \pi^2 [k^2+l^2/r^2]
\end{align}
\item{$\omega=\exp[i \theta]$}\\
The spectrum is characteristic for $T^2/\Zb''_2$ and $T^2/\Db'_2$.
\begin{align}
\label{Mtheta}
M_{k,l} &= \pi \bigl[ k + i ( k \cot \theta - l \sin^{-1} \theta ) \bigr]\\
\label{m2theta}
m^2_{k,l} &= \frac{\pi^2}{\sin^2 \theta} \, \bigl[k^2+l^2-2 k l \cos \theta\bigr] 
\end{align}
\item{$\omega=\exp[i 2 \pi / 3]$}\\
The spectrum is characteristic for $T^2/\Zb_3$, $T^2/\Db_3$ and $T^2/\Fb_3$.
\begin{align}
\label{M2pi3}
M_{k,l} &= \pi \bigl[ k - \frac{i}{\sqrt{3}}(2l+k)\bigr]\\
\label{m22pi3}
m^2_{k,l} &= \frac{4 \pi^2}{3} \, \bigl[k^2 + k l + l^2\bigr]
\end{align}
\item{$\omega= i$}\\
The spectrum is characteristic for $T^2/\Zb_4$, $T^2/\Db_4$ and $\Rb^2/\pIVg$.
\begin{align}
\label{Mi}
M_{k,l} &= \pi [k - i l]\\
\label{m2i}
m^2_{k,l} &= \pi^2 \, [k^2 + l^2]
\end{align}
\item{$\omega=\exp[i \pi / 3]$}\\
The spectrum is characteristic for $T^2/\Zb_6$ and $T^2/\Db_6$.
\begin{align}
\label{Mpi3}
M_{k,l} &= \pi \bigl[ k - \frac{i}{\sqrt{3}}(2l-k)\bigr]\\
\label{m2pi3}
m^2_{k,l} &= \frac{4 \pi^2}{3} \, \bigl[k^2 - k l + l^2\bigr]
\end{align}
\end{itemize}


\subsection{Klein Bottle $\Rb^2/\pg$}
\label{R2pg}

The space group $\pg$ (\ref{pg}) possesses two generators. In the complex plane, we identify them with a translation $t_2$ along the imaginary axis and a glide reflection $g$.
\begin{align}
z &\sim z + i r\\
\label{pg_g}
&\sim z^* + i r + 1/2
\end{align}
The translation $t_1$ along the real axis can be expressed in terms of the glide reflection as $t_1=g^2$. Applying (\ref{pg_g}) twice, we find
\begin{equation}
z \sim z +1 \quad .
\end{equation}
In order to find all possible parities for complex fields $\vp(z)$ on the Klein bottle, we solve the relations (\ref{pg}) in $\Cb$.
\begin{equation}
\label{pgsolutions}
t_2=\pm, \quad g=\exp[i 2 \pi \rho] \qquad \text{with} \quad \rho \in [0,1) \subset \Qb
\end{equation}
A field $\vp(z)$ can therefore be described by one discrete and one continuous parity. Unlike $t_1=g^2=\exp[i 4 \pi \rho]$, the SS phase associated with the translation along the imaginary axis can only be discrete, $t_2=\pm$. Here we ignore SS phases.
\begin{equation}
t_2=+, \quad g=\pm
\end{equation}
A complex field $\vp(z)$ can therefore have only two parities.
\begin{align}
\label{Kb_parity}
\vp(z^*+i r+\frac{1}{2}) &= \pm \vp(z)\\
\vp(z+i r) &= \vp(z)\\
\vp(z+1) &= \vp(z)
\end{align}
The field can be expanded in terms of the following basis functions
\begin{equation}
F^{(\pm)}_{k,l}(z) = c^{(\pm)}_{k,l} \Bigl[ f_{k,l}(z) \pm f_{k,l}(z^*+i r+\frac{1}{2}) \Bigr]
\end{equation}
where $c^{(\pm)}_{k,l}$ are normalization constants and the exponentials $f_{k,l}(z)$ are given by (\ref{fkl}) and (\ref{Mri}). As in the one-dimensional case (\ref{S1sym}), there exists a helpful relation between argument and indices of the exponentials
\begin{equation}
\label{Kbsym}
f_{k,l}(z^* + i r +\frac{1}{2}) = (-)^k f_{k,-l}(z)
\end{equation}
which allows us to simplify the basis functions.
\begin{align}
\label{Kb_Fkl}
F^{(\pm)}_{k,l}(z) &= c^{(\pm)}_{k,l} \bigl[ f_{k,l}(z) \pm (-)^k f_{k,-l}(z) \bigr]\\
\label{Kb_FklII}
&=\sqrt{2^{-1-\delta_{l,0}}} \, \Bigl( \exp \bigl[i 2 \pi (k y_5 + l y_6)\bigr] \pm (-)^k \exp \bigl[i 2 \pi (k y_5 - l y_6)\bigr] \Bigr)
\end{align}
The normalization constants 
\begin{equation}
c^{(\pm)}_{k,l}=\sqrt{2^{-1-\delta_{l,0}}}
\end{equation}
have been determined from the orthonormality condition.
\begin{equation}
\label{Kborth}
\int \dzz \; F^{(p)}_{k,l}(z) F^{(q)*}_{m,n}(z) = \delta_{p,q} \, \delta_{k,m} \, \delta_{l,n}
\end{equation}
The proof of (\ref{Kborth}) relies on the orthonormality of the $f_{k,l}(z)$ on the torus. 
As in (\ref{T2orth}), we integrate in the above expression over the fundamental domain of $T^2$. Note that (\ref{Kborth}) conforms with our discussion of the one-dimensional case, see paragraph below (\ref{S1Z2orth}). Complex fields $\vp(z)$ can have two parities (\ref{Kb_parity}), and we consequently integrate twice over the fundamental domain of the orbifold. The functions (\ref{Kb_FklII}) are periodic in $0 \leq y_{5,6} \leq 1$.
\begin{equation}
\label{Kbwave}
\bigl[ \pz \pzc + m_{k,l}^2\bigr] F^{(\pm)}_{k,l}(z) =0
\end{equation}
The spectrum on the Klein bottle is independ of the parities and given by (\ref{m2ri}). A proof of (\ref{Kbwave}) relies on (\ref{T2wave}) and $m^2_{k,l}=m^2_{k,-l}$. Zero modes can vanish for both parities.
\begin{equation}
F^{(+)}_{2k+1,0}(z)=F^{(-)}_{2k,0}(z)=0
\end{equation}
Relation (\ref{Kbsym}) implies that the basis functions $F^{(\pm)}_{k,l}(z)$ are not independent of each other.
\begin{equation}
\label{KbRestrict}
F^{(\pm)}_{k,l}(z^*+ir+\frac{1}{2}) = (-)^k F^{(\pm)}_{k,-l}(z)=\pm F^{(\pm)}_{k,l}(z)
\end{equation}
We therefore restrict one of the indices in the expansion of the complex fields, $l \geq 0$.
\begin{equation}
\label{Kbexpansion}
\vp(z) = \sum_{\substack{k=-\infty\\ l=0}}^{\infty} \; \vp_{(k,l)} F^{(\pm)}_{k,l}(z)
\end{equation}
Following our discussion in Section~\ref{S1Z2}, we can construct the expansion of the delta function by substituting an even ansatz\footnote{Note that the positive parity implies $\delta(z^*+i r+1/2)=\delta(z) \neq \delta(-z)$ and $\delta(z-z') \neq \delta(z'-z)$.} in the defining relation (\ref{DefDelta2D}) and check the completeness of our basis.
\begin{gather}
\label{DefDelta2D}
\int \dzz \; \vp(z) \delta^*(z-z') = \vp(z')\\
\label{Kbdelta}
\delta(z) = \sum_{\substack{k=-\infty\\ l=0}}^{\infty} \; F^{(+)*}_{k,l}(0) F^{(+)}_{k,l}(z)\\
\label{Kbcomplete}
\delta (z_1-z_2) = \sum_{p=\pm} \; \sum_{\substack{k=-\infty\\ l=0}}^{\infty} \; F^{(p)}_{k,l} (z_1) \, F^{(p)*}_{k,l} (z_2)
\end{gather}


\subsection{Real Projective Plane $\Rb P^2$}
\label{RP2}

We consider the real projective plane\footnote{In~\cite{Hebecker:2003we} the author discusses a 6D grand unified theory compactified on a space topologically equivalent to $\Rb P^2$.} as a quotient space of the real plane $\Rb^2$ modulo the discrete space group $\pgg$ (\ref{pgg}). The group is generated by the $\pi$-rotation $r$ and the glide reflection $g$. In the complex plane, they correspond to the following transformations. 
\begin{align}
\label{pgg_r}
z &\sim -z\\
\label{pgg_g}
&\sim z^* + \frac{i r}{2} + \frac{1}{2}
\end{align}
The two translations $t_1 = g^2$ and $t_2 = [gr]^2$ identify the points
\begin{align}
z &\sim z+1\\
&\sim z+i r \; .
\end{align}
In order to determine all available parities on $\Rb P^2$, we solve the relations (\ref{pgg}) in $\Cb$.
\begin{equation}
r=\pm, \quad g=\pm
\end{equation}
Note that no SS phases are possible on the real projective plane, since $t_1 = g^2 = +$ and $t_2 = [gr]^2 = +$. A complex field $\vp(z)$ is therefore described by four distinct parities $(p,q)$ with $p,q=\pm$.
\begin{align}
\vp(-z) &= p \, \vp(z)\\
\vp(z^*+\frac{i}{2}+\frac{1}{2}) &= q \, \vp(z)\\
\vp(z+1) &= \vp(z)\\
\vp(z+i r) &= \vp(z)
\end{align}
The field can be expanded in terms of basis functions $F^{(p,q)}_{k,l}(z)$. Using (\ref{fkl}), (\ref{Mri}) and
\begin{align}
\label{RP2sym1}
f_{k,l}(-z) &= f_{-k,-l}(z)\\
\label{RP2sym2}
f_{k,l}(z^*+\frac{ir}{2}+\frac{1}{2}) &= (-)^{k+l} \; f_{k,-l}(z)
\end{align}
we find their explicit form.
\begin{align}
\label{RP2_Fkl}
F^{(p,q)}_{k,l}(z) &= \sqrt{2^{-2-\delta_{k,0}-\delta_{l,0}}} \bigl[ f_{k,l}(z) + p \, f_{-k,-l}(z) + (-)^{k+l} q \, f_{k,-l}(z) + (-)^{k+l} p q \, f_{-k,l}(z) \bigr]\\
&= \sqrt{2^{-\delta_{k,0}-\delta_{l,0}}} \; \begin{cases} \cos 2 \pi (k y_5 + l y_6) + (-)^{k+l} q \; \cos 2 \pi (k y_5 - l y_6) \qquad &\text{for} \quad p=+\\ i \sin 2 \pi (k y_5 + l y_6) + (-)^{k+l} q i \; \sin 2 \pi (k y_5 - l y_6) \qquad &\text{for} \quad p=- \end{cases} \nonumber
\end{align}
The normalization constants in the above expression are derived from the orthonormality condition. In (\ref{RP2orth}) we integrate four times over the orbifold fundamental domain, i.e. the torus of worldvolume $r$.
\begin{equation}
\label{RP2orth}
\int \dzz \; F^{(p,q)}_{k,l}(z) F^{(r,s)*}_{m,n}(z) = \delta_{p,r} \, \delta_{q,s} \, \delta_{k,m} \, \delta_{l,n}
\end{equation}
The mass spectrum is given by (\ref{m2ri}), since $m^2_{k,l}=m^2_{-k,-l}=m^2_{k,-l}=m^2_{-k,l}$.
\begin{equation}
\bigl[ \pz \pzc + m_{k,l}^2\bigr] F^{(p,q)}_{k,l}(z) =0
\end{equation}
Zero modes vanish for the following combinations of Kaluza-Klein modes and parities.
\begin{equation}
\label{RP2zeromodes}
F^{(\pm,+)}_{2k+1,0}(z)=F^{(\pm,-)}_{2k,0}(z)=F^{(p,p)}_{0,2k+1}(z)=F^{(p,-p)}_{0,2k}(z)=0 \qquad \text{with} \quad p=\pm
\end{equation}
Using (\ref{RP2sym1}) and (\ref{RP2sym2}), we find that the $F^{(p,q)}_{k,l}(z)$ are not independent.
\begin{gather}
\label{here}
F^{(p,q)}_{k,l} (z^* + \frac{ir}{2} + \frac{1}{2}) = (-)^{k+l} \, F^{(p,q)}_{k,-l}(z) = q \, F^{(p,q)}_{k,l} (z)\\
\label{andhere}
F^{(p,q)}_{k,l} (-z) = F^{(p,q)}_{-k,-l} (z) = p F^{(p,q)}_{k,l} (z)
\end{gather}
The second equalities in (\ref{here}) and (\ref{andhere}) restrict the indices $l$ and $k$ respectively, $l \geq 0$ and $k \geq 0$. A complex field $\vp(z)$ can therefore be expanded as follows.
\begin{equation}
\vp(z) = \sum_{k,l=0}^\infty \; \vp_{(k,l)} F^{(p,q)}_{k,l}(z)
\end{equation}
Following the discussion in Section~\ref{S1Z2}, we can find the expansion of the delta function and check the completeness of our basis.


\subsection{$T^2/\Zb_2$}
\label{T2Z2}

The group $\pII$ (\ref{p2}) is generated by the two translations $t_1$ and $t_2$ as well as the $\pi$-rotation $r$. In the complex plane, we identify them with the following transformations.
\begin{align}
\label{T2Z2def1}
z &\sim z + 1\\
&\sim z + \omega \qquad \text{with} \quad \omega= r \, \exp[i \theta]\\
\label{T2Z2def3}
&\sim -z \; ,
\end{align}
Solving (\ref{p2}) in $\Cb$, we find eight solutions.
\begin{equation}
\label{p2solutions}
t_1=\pm, \quad t_2=\pm, \quad r=\pm
\end{equation}
SS phases on $T^2/\Zb_2$ will be discussed in Appendix~\ref{SSonT2Z2}. Here we will restrict ourselves to
\begin{equation}
t_1=t_2=+ , \qquad r=\pm \; .
\end{equation}
A complex field $\vp(z)$ can therefore have only two different parities. 
\begin{align}
\label{phi1}
\vp(- z) &= \pm \vp(z)\\
\vp(z+1) &= \vp(z)\\
\label{phi3}
\vp(z+\omega) &= \vp(z)
\end{align}
The field can be expanded in terms of basis functions $F^{(\pm)}_{k,l}(z)$. Using (\ref{fkl}), (\ref{M}) and
\begin{equation}
\label{T2Z2sym}
f_{k,l}(-z) = f_{-k,-l}(z)
\end{equation}
we find their explicit form.
\begin{align}
F^{(p)}_{k,l}(z) &= \sqrt{2^{-1-\delta_{k,0} \delta_{l,0}}} \, \bigl[ f_{k,l}(z) + p \; f_{-k,-l}(z) \bigr]\\
&= \begin{cases} \sqrt{2^{1-\delta_{k,0} \delta_{l,0}}} \, \cos 2 \pi (k y_5 + l y_6) \qquad &\text{for} \quad p=+\\ \sqrt{2} i \, \sin 2 \pi (k y_5 + l y_6) \qquad &\text{for} \quad p=- \end{cases}
\end{align}
The normalization constants in the above expression are derived from the orthonormality condition. In (\ref{T2Z2orth}) we integrate twice over the orbifold fundamental domain, that is the torus of worldvolume $r \sin \theta$, see Fig.~3(a).
\begin{equation}
\label{T2Z2orth}
\int \dzz \; F^{(p)}_{k,l}(z) F^{(q)*}_{m,n}(z) = \delta_{p,q} \, \delta_{k,m} \, \delta_{l,n}
\end{equation}
The mass spectrum is given by (\ref{m2}), since $m_{k,l}^2=m_{-k,-l}^2$.
Derivatives relate the different parities to each other.
\begin{equation}
\bigl[ \pz \pzc + m_{k,l}^2\bigr] F^{(p)}_{k,l}(z) =0
\end{equation}
\begin{align}
\pz F^{(\pm)}_{k,l}(z) &= i M_{k,l} \; F^{(\mp)}_{k,l}(z)\\
\pzc F^{(\pm)}_{k,l}(z) &= i M^*_{k,l} \; F^{(\mp)}_{k,l}(z)
\end{align}
There is only a single vanishing zero mode.
\begin{equation}
F^{(-)}_{0,0}(z)=0
\end{equation}
Using (\ref{T2Z2sym}), we find that the $F^{(\pm)}_{k,l}(z)$ are not independent.
\begin{equation}
F^{(\pm)}_{k,l} (-z) = F^{(\pm)}_{-k,-l}(z) = \pm \, F^{(\pm)}_{k,l}(z)
\end{equation}
We restrict the indices to $-\infty < k < \infty$, $l \geq 0$ for $k \geq 0$ and $l \geq 1$ for $k <0$.
Unlike in (\ref{KbRestrict}), a restriction $l \geq 0$ will not do, since $F^{(\pm)}_{-k,0}$ and $F^{(\pm)}_{k,0}$ are not independent. A complex field $\vp(z)$ can therefore be expanded as follows.
\begin{equation}
\vp(z) = \sum_{\substack{k=-\infty\\ l=1 \; \text{for} \; k < 0\\ l=0 \; \text{for} \; k \geq 0}}^{\infty} \; \vp_{(k,l)} F^{(\pm)}_{k,l}(z)
\end{equation}
Following the discussion in Section~\ref{S1Z2}, we can find the expansion of the delta function and check the completeness of our basis.


\subsection{$T^2/\Zb_2'$}
\label{T2Z2p}

The group $\pM \simeq \Zb \times \Db_\infty$ (\ref{pm}) is generated by the two translations $t_1$ and $t_2$ as well as the reflection~$f$. In the complex plane, they are identified with the following transformations.
\begin{align}
\label{pm_t1}
z &\sim z + 1\\
\label{pm_t2}
&\sim z + r i\\
\label{pm_f}
&\sim z^*
\end{align}
In order to determine all available parities on $T^2/\Zb_2'$, we solve the relations (\ref{pm}) in $\Cb$.
\begin{equation}
t_1 = \exp[i 2 \pi \rho], \quad t_2=\pm, \quad f=\pm \qquad \text{with} \quad \rho \in [0,1) \subset \Qb
\end{equation}
SS phases associated with (\ref{pm_t1}) are unrestricted. The $\Db_\infty$ structure of translation (\ref{pm_t2}) and reflection (\ref{pm_f}) gives rise to the same parities as on $S^1/\Zb_2$, see (\ref{Doo_C}). Here we ignore SS phases.
\begin{equation}
t_1=t_2=+, \quad f=\pm
\end{equation}
A complex field $\vp(z)$ can therefore be either even or odd.
\begin{align}
\vp(z^*) &= \pm \vp(z)\\
\vp(z+1) &= \vp(z)\\
\vp(z+r i) &= \vp(z)
\end{align}
The field can be expanded in terms of basis functions $F^{(\pm)}_{k,l}(z)$. Using (\ref{fkl}), (\ref{Mri}) and
\begin{equation}
\label{T2Z2psym}
f_{k,l}(z^*) = f_{k,-l}(z)
\end{equation}
we find their explicit form.
\begin{align}
F^{(\pm)}_{k,l}(z) &= \sqrt{2^{-1-\delta_{l,0}}} \, \bigl[ f_{k,l}(z) \pm f_{k,-l}(z) \bigr]\\
&= \sqrt{2^{-1-\delta_{l,0}}} \, \Bigl[ \exp \bigl[i 2 \pi (k y_5 + l y_6) \bigr] \pm \exp \bigl[i 2 \pi (k y_5 - l y_6) \bigr]\Bigr]
\end{align}
The normalization constants in the above expression are derived from the orthonormality condition. In (\ref{T2Z2porth}) we integrate twice over the orbifold domain, i.e. the torus.
\begin{equation}
\label{T2Z2porth}
\int \dzz \; F^{(p)}_{k,l}(z) F^{(q)*}_{m,n}(z) = \delta_{p,q} \, \delta_{k,m} \, \delta_{l,n}
\end{equation}
The mass spectrum is given by (\ref{m2ri}).
\begin{equation}
\bigl[ \pz \pzc + m_{k,l}^2\bigr] F^{(p)}_{k,l}(z) =0
\end{equation}
Zero modes vanish only for odd basis functions.
\begin{equation}
F^{(-)}_{k,0}(z)=0
\end{equation}
Using (\ref{T2Z2psym}), we find that the $F^{(\pm)}_{k,l}(z)$ are not independent.
\begin{equation}
F^{(\pm)}_{k,l} (z^*) = F^{(\pm)}_{k,-l}(z) = \pm \, F^{(\pm)}_{k,l}(z)
\end{equation}
We restrict the second index to $l \geq 0$ and expand complex fields as follows.
\begin{equation}
\vp(z) = \sum_{\substack{k=-\infty\\ l=0}}^{\infty} \; \vp_{(k,l)} F^{(\pm)}_{k,l}(z)
\end{equation}
Following the discussion in Section~\ref{S1Z2}, we can find the expansion of the delta function and check the completeness of our basis.


\subsection{$T^2/\Zb_2''$}
\label{T2Z2pp}

The group $\cm$ (\ref{cm}) is generated by the two translations $t_1$ and $t_2$ as well as the reflection~$f$. In the complex plane, they correspond to the following transformations.
\begin{align}
z &\sim z + 1\\
&\sim z + \omega \qquad \text{with} \quad \omega= \exp[i \theta]\\
&\sim \omega z^* \; ,
\end{align}
The relations (\ref{cm}) have got the following solutions in $\Cb$.
\begin{equation}
\label{cmSolution}
t_1=t_2=\exp[i 2 \pi \rho], \quad f=\pm \qquad \text{with} \quad \rho \in [0,1) \subset \Qb
\end{equation}
The SS phases are continuous, but need to be identical in both directions. Here we ignore these phases and restrict ourselves to the subgroup $\Zb_2 \subseteq \cm$.
\begin{equation}
t_1=t_2=+, \quad f=\pm
\end{equation}
A complex field can therefore be either even or odd.
\begin{align}
\vp(\omega z^*) &= \pm \vp(z)\\
\vp(z+1) &= \vp(z)\\
\vp(z+\omega) &= \vp(z)
\end{align}
The field can be expanded in terms of basis functions $F^{(\pm)}_{k,l}(z)$. Using (\ref{fkl}), (\ref{Mtheta}) and
\begin{equation}
\label{T2Z2ppsym}
f_{k,l}(\omega z^*) = f_{l,k}(z)
\end{equation}
we find their explicit form.
\begin{align}
F^{(\pm)}_{k,l}(z) &= \sqrt{2^{-1-\delta_{k,l}}} \, \bigl[ f_{k,l}(z) \pm f_{l,k}(z) \bigr]\\
&= \sqrt{2^{-1-\delta_{k,l}}} \, \Bigl[ \exp \bigl[i 2 \pi (k y_5 + l y_6) \bigr] \pm \exp \bigl[i 2 \pi (l y_5 + k y_6) \bigr]\Bigr]
\end{align}
The normalization constants in the above expression are derived from the orthonormality condition. In (\ref{T2Z2pporth}) we integrate twice over the orbifold fundamental domain, i.e. the torus.
\begin{equation}
\label{T2Z2pporth}
\int \dzz \; F^{(p)}_{k,l}(z) F^{(q)*}_{m,n}(z) = \delta_{p,q} \, \delta_{k,m} \, \delta_{l,n}
\end{equation}
The mass spectrum is given by (\ref{m2theta}), since $m_{k,l}^2=m_{l,k}^2$.
\begin{equation}
\bigl[ \pz \pzc + m_{k,l}^2\bigr] F^{(p)}_{k,l}(z) =0
\end{equation}
Only odd zero modes vanish.
\begin{equation}
F^{(-)}_{k,k}(z)=0
\end{equation}
Using (\ref{T2Z2ppsym}), we find that the $F^{(p)}_{k,l}(z)$ are not independent.
\begin{equation}
F^{(\pm)}_{k,l} ( \omega z^*) = F^{(\pm)}_{l,k}(z) = \pm \, F^{(\pm)}_{k,l}(z)
\end{equation}
We restrict the indices to $-\infty < k < \infty$, $l \geq k$ and expand complex fields as follows.
\begin{equation}
\vp(z) = \sum_{\substack{k=-\infty\\ l=k}}^{\infty} \; \vp_{(k,l)} F^{(\pm)}_{k,l}(z)
\end{equation}
Following the discussion in Section~\ref{S1Z2}, we can find the expansion of the delta function and check the completeness of our basis.


\subsection{$T^2/\Db_2$}
\label{T2D2}

The group $\pmm \simeq \Db_\infty^2$ (\ref{pmm}) is generated by the two translations $t_1$ and $t_2$ as well as the $\pi$-rotation~$r$ and the reflection~$f$. They are identified with the following transformations in the complex plane.
\begin{align}
z &\sim z + 1\\
&\sim z + r i\\
\label{pmm_r}
&\sim -z\\
\label{pmm_f}
&\sim z^*
\end{align}
The relations (\ref{pmm}) have got 16 solutions in $\Cb$.
\begin{equation}
t_1=\pm, \quad t_2=\pm, \quad r=\pm, \quad f=\pm 
\end{equation}
Here we ignore SS phases and restrict ourselves to the subgroup $\Db_2 \subseteq \pmm$.
\begin{equation}
t_1=t_2=+, \quad r=\pm, \quad f=\pm
\end{equation}
We consider complex fields $\vp(z)$ with four distinct parities $(p,q)$ with $p,q=\pm$.
\begin{align}
\vp(-z) &= p \, \vp(z)\\
\vp(z^*) &= q \, \vp(z)\\
\vp(z+1) &= \vp(z)\\
\vp(z+r i) &= \vp(z)
\end{align}
The fields can be expanded in terms of basis functions $F^{(p,q)}_{k,l}(z)$. Using (\ref{fkl}), (\ref{Mri}) and
\begin{align}
\label{T2D2sym1}
f_{k,l}(-z) &= f_{-k,-l}(z)\\
\label{T2D2sym2}
f_{k,l}(z^*) &= f_{k,-l}(z)
\end{align}
we find their explicit form.
\begin{align}
F^{(p,q)}_{k,l}(z) &= \sqrt{2^{-2-\delta_{k,0}-\delta_{l,0}}} \, \bigl[ f_{k,l}(z) + p \, f_{-k,-l}(z) + q \, f_{k,-l}(z) + p q \, f_{-k,l}(z) \bigr]\\
&= \sqrt{2^{-\delta_{k,0}-\delta_{l,0}}} \, \begin{cases} \cos 2 \pi (k y_5 + l y_6) + q \, \cos 2 \pi (k y_5 - l y_6) \qquad &\text{for} \quad p=+\\ i \sin 2 \pi (k y_5 + l y_6) + q i \, \sin 2 \pi (k y_5 - l y_6) \qquad &\text{for} \quad p=- \end{cases}
\end{align}
The normalization constants in the above expression are derived from the orthonormality condition. In (\ref{T2D2orth}) we integrate four times over the orbifold domain, i.e. the torus.
\begin{equation}
\label{T2D2orth}
\int \dzz \; F^{(p,q)}_{k,l}(z) F^{(r,s)*}_{m,n}(z) = \delta_{p,r} \, \delta_{q,s} \, \delta_{k,m} \, \delta_{l,n}
\end{equation}
The mass spectrum is given by (\ref{m2ri}).
\begin{equation}
\bigl[ \pz \pzc + m_{k,l}^2\bigr] F^{(p,q)}_{k,l}(z) =0
\end{equation}
The following zero modes vanish.
\begin{equation}
F^{(\pm,-)}_{k,0}(z)=F^{(p,-p)}_{0,l}(z)=0 \qquad \text{with} \quad p=\pm
\end{equation}
Using (\ref{T2D2sym1}) and (\ref{T2D2sym2}), we find that the $F^{(p,q)}_{k,l}(z)$ are not independent.
\begin{gather}
F^{(p,q)}_{k,l} (-z) = F^{(p,q)}_{-k,-l}(z) = p \, F^{(p,q)}_{k,l}(z)\\
F^{(p,q)}_{k,l} (z^*) = F^{(p,q)}_{k,-l}(z) = q \, F^{(p,q)}_{k,l}(z)
\end{gather}
We restrict the indices in the expansion to $k,l \geq 0$.
\begin{equation}
\vp(z) = \sum_{k,l=0}^{\infty} \; \vp_{(k,l)} F^{(p,q)}_{k,l}(z)
\end{equation}
Following the discussion in Section~\ref{S1Z2}, we can find the expansion of the delta function and check the completeness of our basis.


\subsection{$T^2/\Db_2'$}
\label{T2D2p}

The group $\cmm$ (\ref{cmm}) is generated by the two translations $t_1$ and $t_2$ as well as the $\pi$-rotation $r$ and the reflection $f$. In the complex plane, they are identified with the following transformations.
\begin{align}
z &\sim z + 1\\
&\sim z + \omega \qquad \text{with} \quad \omega = \exp[i \theta]\\
\label{cmm_r}
&\sim - z\\
& \sim \omega z^*
\end{align}
Solving (\ref{cmm}) in $\Cb$, we find eight solutions.
\begin{equation}
t_1=t_2=\pm, \quad r=\pm, \quad f=\pm
\end{equation}
The two SS phases must be identical. Here we ignore them and work with the subgroup $\Db_2 \subseteq \cmm$.
\begin{equation}
t_1=t_2=+, \quad r=\pm, \quad f=\pm
\end{equation}
A complex field $\vp(z)$ can therefore have four distinct parities $(p,q)$ with $p,q=\pm$.
\begin{align}
\vp(-z) &= p \, \vp(z)\\
\vp(\omega z^*) &= q \, \vp(z)\\
\vp(z+1) &= \vp(z)\\
\vp(z+\omega) &= \vp(z)
\end{align}
The field can be expanded in terms of basis functions $F^{(p,q)}_{k,l}(z)$. Using (\ref{fkl}), (\ref{Mtheta}) and
\begin{align}
\label{T2D2psym1}
f_{k,l}(-z) &= f_{-k,-l}(z)\\
\label{T2D2psym2}
f_{k,l}(\omega z^*) &= f_{l,k}(z)
\end{align}
we find their explicit form.
\begin{align}
F^{(p,q)}_{k,l}(z) &= \sqrt{2^{-2- \delta_{k,l} - \delta_{k,-l}}} \, \bigl[ f_{k,l}(z) + p \, f_{-k,-l}(z) + q \, f_{l,k}(z) + p q \, f_{-l,-k}(z) \bigr]\\
&= \sqrt{2^{-\delta_{k,0}-\delta_{l,0}}} \, \begin{cases} \cos 2 \pi (k y_5 + l y_6) + q \, \cos 2 \pi (l y_5 + k y_6) \qquad &\text{for} \quad p=+\\ i \sin 2 \pi (k y_5 + l y_6) + q i \, \sin 2 \pi (l y_5 + k y_6) \qquad &\text{for} \quad p=- \end{cases}
\end{align}
The normalization constants in the above expression are derived from the orthonormality condition. In (\ref{T2D2porth}) we integrate four times over the orbifold fundamental domain.
\begin{equation}
\label{T2D2porth}
\int \dzz \; F^{(p,q)}_{k,l}(z) F^{(r,s)*}_{m,n}(z) = \delta_{p,r} \, \delta_{q,s} \, \delta_{k,m} \, \delta_{l,n}
\end{equation}
The mass spectrum is given by (\ref{m2theta}).
\begin{equation}
\bigl[ \pz \pzc + m_{k,l}^2\bigr] F^{(p,q)}_{k,l}(z) =0
\end{equation}
The following zero modes vanish.
\begin{equation}
F^{(\pm,-)}_{k,0}(z)=F^{(\pm,-)}_{0,k}(z)=F^{(-,\pm)}_{0,0}(z)=F^{(\pm,-)}_{k,k}(z)=F^{(\pm,\mp)}_{k,-k}(z)=0
\end{equation}
Using (\ref{T2D2psym1}) and (\ref{T2D2psym2}), we find that the $F^{(p,q)}_{k,l}(z)$ are not independent.
\begin{gather}
F^{(p,q)}_{k,l} (-z) = F^{(p,q)}_{-k,-l}(z) = p \, F^{(p,q)}_{k,l}(z)\\
F^{(p,q)}_{k,l} (\omega z^*) = F^{(p,q)}_{l,k}(z) = q \, F^{(p,q)}_{k,l}(z)
\end{gather}
We restrict the indices to $-\infty < k < \infty$ and $l \geq |k|$.
\begin{equation}
\vp(z) = \sum_{\substack{k=-\infty\\ l=|k|}}^{\infty} \; \vp_{(k,l)} F^{(p,q)}_{k,l}(z)
\end{equation}
Following the discussion in Section~\ref{S1Z2}, we can find the expansion of the delta function and check the completeness of our basis.


\subsection{$T^2/\Fb_2$}
\label{T2F2}

The group $\pmg$ (\ref{pmg}) is generated by the two translations $t_1$ and $t_2$ as well as the $\pi$-rotation $r$ and the reflection $f$. In the complex plane, they are identified with the following transformations.
\begin{align}
z &\sim z + 1\\
&\sim z + r i\\
&\sim -z\\
&\sim z^* + \frac{i r}{2}
\end{align}
Solving the relations (\ref{pmg}) in $\Cb$, we find 16 solutions.
\begin{equation}
t_1=\pm, \quad t_2=\pm, \quad r=\pm, \quad f=\pm
\end{equation}
Here we ignore SS phases and restrict ourselves to the subgroup $\Fb_2 \subseteq \pmg$.
\begin{equation}
t_1=t_2=+ , \quad r=\pm , \quad f=\pm
\end{equation}
A complex field $\vp(z)$ can have four distinct parities $(p,q)$ with $p,q=\pm$.
\begin{align}
\vp(-z) &= p \, \vp(z)\\
\vp(z^* + \frac{i r}{2}) &= q \, \vp(z)\\
\vp(z+1) &= \vp(z)\\
\vp(z+r i) &= \vp(z)
\end{align}
The field can be expanded in terms of basis functions $F^{(p,q)}_{k,l}(z)$. Using (\ref{fkl}), (\ref{Mri}) and
\begin{align}
\label{T2F2sym1}
f_{k,l}(-z) &= f_{-k,-l}(z)\\
\label{T2F2sym2}
f_{k,l}(z^* + \frac{i r}{2}) &= (-)^l \, f_{k,-l}(z)
\end{align}
we find their explicit form.
\begin{align}
F^{(p,q)}_{k,l}(z) &= \sqrt{2^{-2-\delta_{k,0}-\delta_{l,0}}} \, \bigl[ f_{k,l}(z) + p \, f_{-k,-l}(z) + (-)^l q \, f_{k,-l}(z) + (-)^l p q \, f_{-k,l}(z) \bigr]\\
&= \sqrt{2^{-\delta_{k,0}-\delta_{l,0}}} \, \begin{cases} \cos 2 \pi (k y_5 + l y_6) + (-)^l q \, \cos 2 \pi (k y_5 - l y_6) \qquad &\text{for} \quad p=+\\ i \sin 2 \pi (k y_5 + l y_6) + (-)^l q i \, \sin 2 \pi (k y_5 - l y_6) \qquad &\text{for} \quad p=- \end{cases} \nonumber
\end{align}
The normalization constants in the above expression are derived from the orthonormality condition. In (\ref{T2F2orth}) we integrate four times over the orbifold fundamental domain.
\begin{equation}
\label{T2F2orth}
\int \dzz \; F^{(p,q)}_{k,l}(z) F^{(r,s)*}_{m,n}(z) = \delta_{p,r} \, \delta_{q,s} \, \delta_{k,m} \, \delta_{l,n}
\end{equation}
The mass spectrum is given by (\ref{m2ri}).
\begin{equation}
\bigl[ \pz \pzc + m_{k,l}^2\bigr] F^{(p,q)}_{k,l}(z) =0
\end{equation}
The following zero modes vanish.
\begin{equation}
F^{(-,+)}_{0,0} = F^{(p,-)}_{0,0}(z) = F^{(p,p)}_{0,2l+1}(z) = F^{(p,-p)}_{0,2l} = F^{(p,-)}_{k,0} =0 \qquad \text{with} \quad p=\pm
\end{equation}
Using (\ref{T2F2sym1}) and (\ref{T2F2sym1}), we find that the $F^{(p,q)}_{k,l}(z)$ are not independent.
\begin{gather}
F^{(p,q)}_{k,l} (-z) = F^{(p,q)}_{-k,-l}(z) = p \, F^{(p,q)}_{k,l}(z)\\
F^{(p,q)}_{k,l} (z^* + \frac{i r}{2}) = (-)^l \, F^{(p,q)}_{k,-l}(z) = q \, F^{(p,q)}_{k,l}(z)
\end{gather}
We restrict the indices to $k,l \geq 0$.
\begin{equation}
\vp(z) = \sum_{k,l=0}^{\infty} \; \vp_{(k,l)} F^{(p,q)}_{k,l}(z)
\end{equation}
Following the discussion in Section~\ref{S1Z2}, we can find the expansion of the delta function and check the completeness of our basis.


\subsection{$T^2/\Zb_3$}
\label{T2Z3}

The group $\pIII$ (\ref{p3}) is generated by the two translations $t_1$ and $t_2$ and the $2 \pi/3$-rotation $r$. In the complex plane, they are identified with the following transformations.
\begin{align}
z &\sim z + 1\\
&\sim z + \omega \qquad \text{with} \quad \omega=\exp[i 2 \pi/3]\\
&\sim \omega z
\end{align}
Solving the relations (\ref{p3}), we find 9 solutions.
\begin{equation}
\label{p3solutions}
t_1=t_2 \in \{+,\omega,\omega^2\}, \quad r \in \{+,\omega,\omega^2\}
\end{equation}
Here we ignore SS phases and restrict ourselves to the subgroup $\Zb_3 \subseteq \pIII$.
\begin{equation}
t_1=t_2=+, \quad r \in \{+,\omega,\omega^2\}
\end{equation}
A complex field $\vp(z)$ can have three different parities.
\begin{align}
\vp(\omega z)&= \omega^n \vp(z)\\
\vp(z+1)&= \vp(z)\\
\vp(z+\omega)&= \vp(z) \, ,
\end{align}
The field can be expanded in terms of $F^{(p)}_{k,l}(z)$ with $p=+,\omega,\omega^2$ for $n=0,1,2$ respectively. Using (\ref{fkl}), (\ref{M2pi3}) and
\begin{equation}
\label{T2Z3sym}
f_{k,l}(\omega z) = f_{l,-k-l}(z)
\end{equation}
we find their explicit form.
\begin{equation}
\label{T2Z3F}
F^{(p)}_{k,l}(z) = \sqrt{3^{-1-\delta_{k,0} \delta_{l,0}}} \bigl[ f_{k,l}(z) + p^2 \, f_{l,-k-l}(z) + p \, f_{-k-l,k}(z) \bigr]
\end{equation}
The normalization constants in the above expression are derived from the orthonormality condition. In (\ref{T2Z3orth}) we integrate three times over the orbifold fundamental domain.
\begin{equation}
\label{T2Z3orth}
\int \dzz \; F^{(p)}_{k,l}(z) F^{(q)*}_{m,n}(z) = \delta_{p,q} \, \delta_{k,m} \, \delta_{l,n}
\end{equation}
Note that the transformation of the indices in (\ref{T2Z3sym}) is cyclic, in accordance with $\omega^3 \, z = z$.
\begin{equation}
(k,l) \; \rightarrow \; (l,-k-l) \; \rightarrow \; (-k-l,k) \; \rightarrow \; (k,l)
\end{equation}
The mass spectrum is given by (\ref{m22pi3}), since $m_{k,l}^2=m_{l,-k-l}^2=m_{-k-l,k}^2$. Derivatives relate the basis function of different parity.
\begin{equation}
\label{T2Z3wave}
\bigl[ \pz \pzc + m_{k,l}^2\bigr] F^{(p)}_{k,l}(z) =0
\end{equation}
\begin{align}
\pz F^{(p)}_{k,l}(z) &= i M_{k,l} \; F^{(\omega^* p)}_{k,l}(z)\\
\pzc F^{(p)}_{k,l}(z) &= i M^*_{k,l} \; F^{(\omega p)}_{k,l}(z)
\end{align}
There are merely two zero modes vanish.
\begin{equation}
F^{(\omega)}_{0,0}(z) = F^{(\omega^2)}_{0,0}(z) = 0
\end{equation}
Using (\ref{T2Z3sym}), we find that the $F^{(p)}_{k,l}(z)$ are not independent.
\begin{gather}
\label{first}
F^{(p)}_{k,l} (\omega z) = F^{(p)}_{l,-k-l}(z) = p \, F^{(p)}_{k,l}(z)\\
\label{second}
F^{(p)}_{k,l} (\omega^2 z) = F^{(p)}_{-k-l,k}(z) = p^2 \, F^{(p)}_{k,l}(z)
\end{gather}
We restrict the indices to $k,l \geq 0$ and $k,l \leq -1$. A complex field $\vp(z)$ can therfore be expanded as follows.
\begin{equation}
\vp(z) = \Bigl[ \sum_{k,l=-\infty}^{-1} + \sum_{k,l=0}^\infty \; \Bigr] \; \vp_{(k,l)} F^{(p)}_{k,l}(z)
\end{equation}
Following the discussion in Section~\ref{S1Z2}, we can find the expansion of the delta function and check the completeness of our basis.


\subsection{$T^2/\Db_3$}
\label{T2D3}

The group $\pIIImI$ (\ref{p3m1}) is generated by the two translations $t_1$ and $t_2$ as well as the $2 \pi/3$-rotation~$r$ and the reflection~$f$. In the complex plane, we identify them with the following transformations.
\begin{align}
z &\sim z + 1\\
&\sim z + \omega \qquad \text{with} \quad \omega=\exp[i 2 \pi/3]\\
&\sim \omega z\\
&\sim  \omega z^*
\end{align}
Solving (\ref{p3m1}) in $\Cb$, we find 6 solutions.
\begin{equation}
\label{T2D3solution}
t_1=t_2 \in \{+,\omega,\omega^2\}, \quad r=+, \quad f=\pm
\end{equation}
Note the solution $r=+$ is trivial, and (\ref{T2D3solution}) is indeed a representation of $\cm \subseteq \pIIImI$ with $\theta=2 \pi /3$, see (\ref{cmSolution}). There is no non-trivial representation of $\pIIImI$ in $\Cb$. Ignoring SS phases, we restrict ourselves to an even smaller subgroup $\Zb_2 \subseteq \cm \subseteq \pIIImI$, i.e. $t_1=t_2=r=+$ and $f=\pm$. Our discussion reduces to the one of Section \ref{T2Z2pp} with $\theta=2 \pi /3$.


\subsection{$T^2/\Fb_3$}
\label{T2F3}

The group $\pIIIIm$ (\ref{p31m}) is generated by the two translations $t_1$ and $t_2$ as well as the $2 \pi/3$-rotation $r$ and the reflection $f$. In the complex plane, we identify them with the following transformations.
\begin{align}
z &\sim z + 1\\
&\sim z + \omega \qquad \text{with} \quad \omega=\exp[i 2 \pi/3]\\
&\sim \omega z\\
&\sim - \omega z^* - \omega^2
\end{align}
The relations (\ref{p31m}) have got 18 solutions in $\Cb$.
\begin{equation}
t_1=t_2 \in \{+,\omega,\omega^2\}, \quad r \in \{+,\omega,\omega^2\}, \quad f=\pm
\end{equation}
Here we ignore SS phases and restrict ourselves to the subgroup $\Fb_3 \subseteq \pIIIIm$.
\begin{equation}
t_1=t_2=+, \quad r \in \{+,\omega,\omega^2\}, \quad f=\pm
\end{equation}
A complex field $\vp(z)$ can therefore posses six different parities.
\begin{align}
\vp(\omega z) &= \omega^n \vp(z)\\
\vp(- \omega z^* - \omega^2) &= \pm \vp(z)\\
\vp(z+1)&= \vp(z)\\
\vp(z+\omega)&= \vp(z)
\end{align}
The field can be expanded in terms of basis functions $F_{k,l}^{(p,q)}(z)$ with $p=+,\omega,\omega^2$ for $n=0,1,2$ respectively, and $q= \pm$. Using (\ref{fkl}), (\ref{M2pi3}) and
\begin{align}
\label{T2F3sym1}
f_{k,l}(\omega z) &= f_{l,-k-l}(z)\\
\label{T2F3sym2}
f_{k,l}(-\omega z^*-\omega^2) &= f_{-l,-k}(z)
\end{align}
we find their explicit form.
\begin{align}
\begin{split}
F^{(p,q)}_{k,l}(z) &= c^{(p,q)}_{k,l} \, \bigl[ f_{k,l}(z) + p^2 f_{l,-k-l}(z) + p f_{-k-l,k}(z)\\
&\qquad + q f_{-l,-k}(z) + p^2 q f_{k+l,-l}(z) + pq f_{-k,k+l}(z) \bigr]
\end{split}\\
c^{(p,q)}_{k,l} &=\sqrt{2^{-1-\delta_{k,0}-\delta_{l,0}+\delta_{k,0}\delta_{l,0}} \; 3^{-1-\delta_{k,0}\delta_{l,0}}}
\end{align}
The normalization constants in the above expression are derived from the orthonormality condition. In (\ref{T2F3orth}) we integrate six times over the orbifold fundamental domain, i.e. the torus.
\begin{equation}
\label{T2F3orth}
\int \dzz \; F^{(p,q)}_{k,l}(z) F^{(r,s)*}_{m,n}(z) = \delta_{p,r} \, \delta_{q,s}\, \delta_{k,m} \, \delta_{l,n}
\end{equation}
Note that the transformation of indices in (\ref{T2F3sym1}) and (\ref{T2F3sym2}) are again cyclic. The mass spectrum is given by (\ref{m22pi3}).
\begin{equation}
\bigl[ \pz \pzc + m_{k,l}^2\bigr] F^{(p,q)}_{k,l}(z) =0
\end{equation}
The following zero modes vanish.
\begin{equation}
F^{(+,-)}_{0,0}=F^{(\omega,\pm)}_{0,0}=F^{(\omega^2,\pm)}_{0,0}=F^{(+,-)}_{0,k}=F^{(+,-)}_{k,0}=0
\end{equation}
Using (\ref{T2F3sym1}) and (\ref{T2F3sym2}), we find that the $F^{(p,q)}_{k,l} (z)$ are not independent.
\begin{gather}
F^{(p,q)}_{k,l} (\omega z) = F^{(p,q)}_{l,-k-l}(z) = p \, F^{(p,q)}_{k,l}(z)\\
F^{(p,q)}_{k,l} (\omega^2 z) = F^{(p,q)}_{-k-l,k}(z) = p^2 \, F^{(p,q)}_{k,l}(z)\\
F^{(p,q)}_{k,l} (-\omega z^*-\omega^2) = F^{(p,q)}_{-l,-k}(z) = q \, F^{(p,q)}_{k,l}(z)
\end{gather}
We restrict the indices in the expansion to $k,l \geq 0$.
\begin{equation}
\vp(z) = \sum_{k,l=0}^\infty \; \vp_{(k,l)} F^{(p,q)}_{k,l}(z)
\end{equation}
Following the discussion in Section~\ref{S1Z2}, we can find the expansion of the delta function and check the completeness of our basis.


\subsection{$T^2/\Zb_4$}
\label{T2Z4}

The group $\pIV$ (\ref{p4}) is generated by the two translations $t_1$ and $t_2$ as well as the $\pi/2$-rotation $r$. In the complex plane, we identify them with the following transformations.
\begin{align}
z &\sim z + 1\\
&\sim z + i\\
&\sim i z
\end{align}
Solving the relations (\ref{p4}) in $\Cb$, we find 16 solutions.
\begin{equation}
t_1=t_2 \in \{+, \, i, \, -, \, -i\}, \qquad r \in \{+, \, i, \, -, \, -i\}
\end{equation}
Here we ignore SS phases and restrict ourselves to the subgroup $\Zb_4 \subseteq \pIV$.
\begin{equation}
t_1=t_2=+, \qquad r \in \{+, \, i, \, -, \, -i\}
\end{equation}
A complex field $\vp(z)$ can therefore posses four parities.
\begin{align}
\vp(iz) &= i^{n} \vp(z)\\
\vp(z+1) &= \vp(z)\\
\vp(z+i) &= \vp(z)
\end{align}
The field can be expanded in terms of $F^{(p)}_{k,l}(z)$ with $p=+,i,-,-i$ for $n=0,1,2,3$ respectively. Using (\ref{fkl}), (\ref{Mi}) and
\begin{equation}
\label{T2Z4sym}
f_{k,l}(iz) = f_{l,-k}(z)
\end{equation}
we find their explicit form.
\begin{align}
F^{(p)}_{k,l}(z) &= 2^{-1-\delta_{k,0} \delta_{l,0}} \, \bigl[ f_{k,l}(z) + p^3 f_{l,-k}(z) + p^2 f_{-k,-l}(z) + p f_{-l,k}(z) \bigr]\\
&=\begin{cases} 2^{-\delta_{k,0} \delta_{l,0}} \Bigl[ \cos 2 \pi (k y_5 + l y_6) + \cos 2 \pi (l y_5 - k y_6) \Bigr] \qquad &\text{for} \quad p=+\\
i \sin 2 \pi (k y_5 + l y_6) + \sin 2 \pi (l y_5 - k y_6) \qquad &\text{for} \quad p=i\\
\cos 2 \pi (k y_5 + l y_6) - \cos 2 \pi (l y_5 - k y_6) \qquad &\text{for} \quad p=-\\
i \sin 2 \pi (k y_5 + l y_6) - \sin 2 \pi (l y_5 - k y_6) \qquad &\text{for} \quad p=-i
\end{cases}
\end{align}
The normalization constants in the above expression are derived from the orthonormality condition. In (\ref{T2Z4orth}) we integrate four times over the orbifold fundamental domain.
\begin{equation}
\label{T2Z4orth}
\int \dzz \; F^{(p)}_{k,l}(z) F^{(q)*}_{m,n}(z) = \delta_{p,q} \, \delta_{k,m} \, \delta_{l,n}
\end{equation}
The transformation of the indices in (\ref{T2Z3sym}) is cyclic, in accordance with $i^4 \, z = z$.
\begin{equation}
(k,l) \; \rightarrow \; (l,-k) \; \rightarrow \; (-k,-l) \; \rightarrow \; (-l,k) \; \rightarrow \; (k,l)
\end{equation}
The mass spectrum is given by (\ref{m2i}), since $m_{k,l}^2=m_{l,-k}^2=m^2_{-k,-l}=m^2_{-l,k}$. Derivatives relate the different parities.
\begin{equation}
\bigl[ \pz \pzc + m_{k,l}^2\bigr] F^{(p)}_{k,l}(z) =0
\end{equation}
\begin{align}
\pz F^{(p)}_{k,l}(z) &= i M_{k,l} \; F^{(-ip)}_{k,l}(z)\\
\pzc F^{(p)}_{k,l}(z) &= i M^*_{k,l} \; F^{(ip)}_{k,l}(z)
\end{align}
The following zero modes vanish.
\begin{equation}
F^{(p)}_{0,0}(z) = 0 \qquad \text{with} \quad p \in \{i,\, -, \, -i\}
\end{equation}
Using (\ref{T2Z4sym}), we find that the $F^{(p)}_{k,l}(z)$ are not independent.
\begin{gather}
F^{(p)}_{k,l} (iz) = F^{(p)}_{l,-k}(z) = p \, F^{(p)}_{k,l}(z)\\
F^{(p)}_{k,l} (-z) = F^{(p)}_{-k,-l}(z) = p^2 \, F^{(p)}_{k,l}(z)
\end{gather}
We restrict the indices to $k,l \geq 0$. A complex field $\vp(z)$ can therefore be expanded as follows.
\begin{equation}
\vp(z) = \sum_{k,l=0}^\infty \; \vp_{(k,l)} F^{(p)}_{k,l}(z)
\end{equation}
Following the discussion in Section~\ref{S1Z2}, we can find the expansion of the delta function and check the completeness of our basis. As discussed in~\cite{Dobrescu:2004zi}, compactification on $T^2/\Zb_4$ is equivalent to compactification on the Chiral Square.


\subsection{$T^2/\Db_4$}
\label{T2D4}

The group $\pIVm$ (\ref{p4m}) is generated by the two translations $t_1$ and $t_2$ as well as the $\pi/2$-rotation $r$ and the reflection $f$. In the complex plane, we identify them with the following transformations.
\begin{align}
z &\sim z + 1\\
&\sim z + i\\
&\sim i z\\
&\sim i z^*
\end{align}
Solving (\ref{p4m}) in $\Cb$, we find 16 solutions.
\begin{equation}
t_1=t_2 \in \{+, \, i, \, -, \, -i\}, \qquad r=\pm, \qquad f=\pm
\end{equation}
Here we ignore SS phases and restrict ourselves to the subgroup $\Db_2 \subseteq \Db_4 \subseteq \pIVm$. Note that there is no non-trivial representation of $\Db_4$ in $\Cb$.
\begin{equation}
t_1=t_2=+, \qquad r=\pm, \qquad f=\pm
\end{equation}
A complex field $\vp(z)$ can therefore posses four parities.
\begin{align}
\vp(iz) &= \pm \vp(z)\\
\vp(iz^*) &= \pm \vp(z)\\
\vp(z+1) &= \vp(z)\\
\vp(z+i) &= \vp(z)
\end{align}
The field can be expanded in terms of the basis $F^{(p,q)}_{k,l}(z)$ with $p,q=\pm$. Using (\ref{fkl}), (\ref{Mi}) and
\begin{align}
\label{T2D4sym1}
f_{k,l}(iz) &= f_{l,-k}(z)\\
\label{T2D4sym2}
f_{k,l}(iz^*) &= f_{l,k}(z)
\end{align}
we find their explicit form.
\begin{align}
\begin{split}
F^{(p,q)}_{k,l}(z) &= \sqrt{2^{-3-\delta_{k,0}-\delta_{l,0}-\delta_{k,l}}} \, \Bigl( f_{k,l}(z) + p f_{l,-k}(z) + f_{-k,-l}(z) + p f_{-l,k}(z) \\
&\qquad +q \bigl[ f_{l,k}(z) + p f_{-k,l}(z) + f_{-l,-k}(z) + p f_{k,-l}(z) \bigr] \Bigr)
\end{split}\\
&=\begin{cases} \sqrt{2^{-1-\delta_{k,0}-\delta_{l,0}-\delta_{k,l}}} \, \Bigl( \cos 2 \pi (k y_5 + l y_6) + \cos 2 \pi (l y_5 - k y_6)\\
\quad +q \bigl[ \cos 2 \pi (l y_5 + k y_6) + \cos 2 \pi (k y_5 - l y_6) \bigr] \Bigr) \qquad &\text{for} \quad p=+\\
\sqrt{2^{-1-\delta_{k,l}}} \, \Bigl( \cos 2 \pi (k y_5 + l y_6) - \cos 2 \pi (l y_5 - k y_6)\\
\quad + q \bigl[ \cos 2 \pi (l y_5 + k y_6) - \cos 2 \pi (k y_5 - l y_6) \bigr] \Bigr) \qquad &\text{for} \quad p=-
\end{cases}
\end{align}
The normalization constants in the above expression are derived from the orthonormality condition. In (\ref{T2D4orth}) we integrate over the domain of the torus.
\begin{equation}
\label{T2D4orth}
\int \dzz \; F^{(p,q)}_{k,l}(z) F^{(r,s)*}_{m,n}(z) = \delta_{p,q} \, \delta_{r,s} \, \delta_{k,m} \, \delta_{l,n}
\end{equation}
Note that the transformations of the indices in (\ref{T2D4sym1}) and (\ref{T2D4sym2}) are cyclic. The mass spectrum is given by (\ref{m2i}).
\begin{equation}
\bigl[ \pz \pzc + m_{k,l}^2\bigr] F^{(p,q)}_{k,l}(z) =0
\end{equation}
The following zero modes vanish.
\begin{equation}
F^{(+,-)}_{0,k}(z)=F^{(-,\pm)}_{0,k}(z)=F^{(+,-)}_{k,0}(z)=F^{(-,\pm)}_{k,0}(z)=F^{(\pm,-)}_{k,k}(z)=0
\end{equation}
Using (\ref{T2D4sym1}) and (\ref{T2D4sym2}), we find that the $F^{(p,q)}_{k,l}(z)$ are not independent.
We restrict the indices $l \geq k \geq 0$.
\begin{gather}
F^{(p,q)}_{k,l} (iz) = F^{(p,q)}_{l,-k}(z) = p \, F^{(p,q)}_{k,l}(z)\\
F^{(p,q)}_{k,l} (-z) = F^{(p,q)}_{-k,-l}(z) = F^{(p,q)}_{k,l}(z)\\
F^{(p,q)}_{k,l} (iz^*) = F^{(p,q)}_{l,k}(z) = q \, F^{(p,q)}_{k,l}(z)
\end{gather}
We restrict the indices to $l \geq k \geq 0$ and expand complex fields as follows.
\begin{equation}
\vp(z) = \sum_{\substack{k=0\\ l=k}}^\infty \; \vp_{(k,l)} F^{(p,q)}_{k,l}(z)
\end{equation}
Following the discussion in Section~\ref{S1Z2}, we can find the expansion of the delta function and check the completeness of our basis.


\subsection{$\Rb^2/\pIVg$}
\label{R2p4g}

The group $\pIVg$ (\ref{p4g}) is generated by the $\pi/2$-rotation $r$ and the reflection $f$. In the complex plane, we identified them with the following transformations.
\begin{align}
z &\sim i z\\
&\sim - i z^* + \frac{1+i}{2}
\end{align}
The two translations $t_1 = [fr^3]^2$ and $t_2 = [fr]^2$ identify the points
\begin{align}
z &\sim z + 1\\
&\sim z + i \; .
\end{align}
The relations (\ref{p4g}) have got eight solutions in $\Cb$.
\begin{equation}
r \in \{+, \, i, \, -, \, -i\}, \qquad f=\pm
\end{equation}
Consequently only discrete SS phases $t_1 =t_2 = r^2 = \pm$ are possible on $\Rb^2/\pIVg$. Here we ignore these phases $t_1 =t_2 = +$ and restrict ourselves to the subgroup $\Db_2 \subseteq \pIVg$.
\begin{equation}
r = \pm, \qquad f=\pm
\end{equation}
A complex field $\vp(z)$ can posses four distinct parities.
\begin{align}
\vp(iz) &= \pm \vp(z)\\
\vp(- i z^* + \frac{1+i}{2}) &= \pm \vp(z)\\
\vp(z+1) &= \vp(z)\\
\vp(z+i) &= \vp(z)
\end{align}
The field can be expanded in terms of basis functions $F^{(p,q)}_{k,l}(z)$. Using (\ref{fkl}), (\ref{Mi}) and
\begin{align}
\label{R2p4gsym1}
f_{k,l}(iz) &= f_{l,-k}(z)\\
\label{R2p4gsym2}
f_{k,l}(-i z^* + \frac{1+i}{2}) &=  (-)^{k+l} \, f_{-l,-k}(z)
\end{align}
we find their explicit form.
\begin{align}
\begin{split}
F^{(p,q)}_{k,l}(z) &= \sqrt{2^{-3-\delta_{k,0}-\delta_{l,0}-\delta_{k,l}}} \, \Bigl( f_{k,l}(z) + p f_{l,-k}(z) + f_{-k,-l}(z) + p f_{-l,k}(z) \\
&\qquad+(-)^{k+l} q \bigl[ f_{-l,-k}(z) + p f_{k,-l}(z) + f_{l,k}(z) + p f_{-k,l}(z) \bigr] \Bigr)
\end{split}\\
&=\begin{cases} \sqrt{2^{-1-\delta_{k,0}-\delta_{l,0}-\delta_{k,l}}} \, \Bigl( \cos 2 \pi (k y_5 + l y_6) + \cos 2 \pi (l y_5 - k y_6)\\
\quad +(-)^{k+l} q \bigl[ \cos 2 \pi (l y_5 + k y_6) + \cos 2 \pi (k y_5 - l y_6) \bigr] \Bigr) \qquad &\text{for} \quad p=+\\
\sqrt{2^{-1-\delta_{k,l}}} \, \Bigl( \cos 2 \pi (k y_5 + l y_6) - \cos 2 \pi (l y_5 - k y_6)\\
\quad + (-)^{k+l} q \bigl[ \cos 2 \pi (l y_5 + k y_6) - \cos 2 \pi (k y_5 - l y_6) \bigr] \Bigr) \qquad &\text{for} \quad p=-
\end{cases}
\end{align}
The normalization constants in the above expression are derived from the orthonormality condition. In (\ref{R2p4gorth}) we integrate over the domain of the torus.
\begin{equation}
\label{R2p4gorth}
\int \dzz \; F^{(p,q)}_{k,l}(z) F^{(r,s)*}_{m,n}(z) = \delta_{p,q} \, \delta_{r,s} \, \delta_{k,m} \, \delta_{l,n}
\end{equation}
Note that the transformations of the indices in (\ref{R2p4gsym1}) and (\ref{R2p4gsym2}) are cyclic. The mass spectrum is given by (\ref{m2i}).
\begin{equation}
\bigl[ \pz \pzc + m_{k,l}^2\bigr] F^{(p,q)}_{k,l}(z) =0
\end{equation}
The following zero modes vanish.
\begin{equation}
F^{(p,-p)}_{0,2k}(z)=F^{(p,p)}_{0,2k+1}(z)=F^{(p,-p)}_{2k,0}(z)=F^{(p,p)}_{2k+1,0}(z)=F^{(p,-)}_{k,k}(z)=0 \qquad \text{with} \quad p=\pm
\end{equation}
Using (\ref{R2p4gsym1}) and (\ref{R2p4gsym2}), we find that the $F^{(p,q)}_{k,l}(z)$ are not independent.
\begin{gather}
F^{(p,q)}_{k,l} (iz) = F^{(p,q)}_{l,-k}(z) = p \, F^{(p,q)}_{k,l}(z)\\
F^{(p,q)}_{k,l} (-z) = F^{(p,q)}_{-k,-l}(z) = F^{(p,q)}_{k,l}(z)\\
F^{(p,q)}_{k,l} (-i z^* + \frac{1+i}{2}) = (-)^{k+l} \, F^{(p,q)}_{-l,-k}(z) = q \, F^{(p,q)}_{k,l}(z)
\end{gather}
We restrict the indices to $l \geq k \geq 0$ and expand complex fields as follows.
\begin{equation}
\vp(z) = \sum_{\substack{k=0\\ l=k}}^\infty \; \vp_{(k,l)} F^{(p,q)}_{k,l}(z)
\end{equation}
Following the discussion in Section~\ref{S1Z2}, we can find the expansion of the delta function and check the completeness of our basis.


\subsection{$T^2/\Zb_6$}
\label{T2Z6}

The group $\pVI$ (\ref{p6}) is generated by the two translations $t_1$ and $t_2$ as well as the $\pi/3$-rotation $r$. In the complex plane, we identified them with the following transformations.
\begin{align}
z &\sim z + 1\\
&\sim z + \omega \qquad \text{with} \quad \omega=\exp[i \pi/3]\\
&\sim \omega z
\end{align}
Solving the relations in (\ref{p6}) in $\Cb$, we find six solutions.
\begin{equation}
\label{p6solutions}
t_1=t_2=+, \qquad r \in \{ +, \omega, \ldots, \omega^5 \}
\end{equation}
Note that it is impossible to assign SS phases to complex fields on $T^2/\Zb_6$. There is no non-trivial representation of $\pVI$ in $\Cb$. A complex field $\vp(z)$ can posses six parities.
\begin{align}
\vp(\omega z) &= \omega^{n} \vp(z)\\
\vp(z+1) &= \vp(z)\\
\vp(z+\omega) &= \vp(z)
\end{align}
The field can be expanded in terms of $F^{(p)}_{k,l}(z)$. Using (\ref{fkl}), (\ref{Mpi3}) and
\begin{equation}
\label{T2Z6sym}
f_{k,l}(\omega z) = f_{l,l-k}(z)
\end{equation}
we find their explicit form.
\begin{equation}
\begin{split}
F^{(p)}_{k,l}(z) = \sqrt{6^{-1-\delta_{k,0} \delta_{l,0}}} \, \bigl[ &f_{k,l}(z) + p^5 f_{l,l-k}(z) + p^4 f_{l-k,-k}(z)\\
+ p^3 &f_{-k,-l}(z) + p^2 f_{-l,k-l}(z) + p f_{k-l,k}(z) \bigr]
\end{split}
\end{equation}
The normalization constants in the above expression are derived from the orthonormality condition. In (\ref{T2Z6orth}) we integrate six times over the fundamental orbifold domain, i.e. the torus.
\begin{equation}
\label{T2Z6orth}
\int \dzz \; F^{(p)}_{k,l}(z) F^{(q)*}_{m,n}(z) = \delta_{p,q} \, \delta_{k,m} \, \delta_{l,n}
\end{equation}
Note that the transformation of the indices in (\ref{T2Z6sym}) is cyclic.
\begin{equation}
(k,l) \; \rightarrow \; (l,l-k) \; \rightarrow \; (l-k,-k) \; \rightarrow \; (-k,-l) \; \rightarrow \; (-l,k-l) \; \rightarrow \; (k-l,k) \; \rightarrow \; (k,l)
\end{equation}
The mass spectrum is given by (\ref{m2pi3}), since $m_{k,l}^2=m_{l,l-k}^2=\cdots$. Derivatives relate the basis functions of different parities.
\begin{gather}
\bigl[ \pz \pzc + m_{k,l}^2\bigr] F^{(p)}_{k,l}(z) =0
\end{gather}
\begin{align}
\pz F^{(p)}_{k,l}(z) &= i M_{k,l} \; F^{(\omega^* p)}_{k,l}(z)\\
\pzc F^{(p)}_{k,l}(z) &= i M^*_{k,l} \; F^{(\omega p)}_{k,l}(z)
\end{align}
The following zero modes vanish.
\begin{equation}
F^{(p)}_{0,0}(z)=0 \qquad \text{with} \quad p\in \{ \omega, \ldots, \omega^5 \}
\end{equation}
Using (\ref{T2Z6sym}), we find that the $F^{(p)}_{k,l}(z)$ are not independent.
\begin{gather}
F^{(p)}_{k,l} (\omega z) = F^{(p)}_{l,l-k}(z) = p \, F^{(p)}_{k,l}(z)\\
F^{(p)}_{k,l} (\omega^2 z) = F^{(p)}_{l-k,-k}(z) = p^2 \, F^{(p)}_{k,l}(z)
\end{gather}
We restrict the indices to $l \geq k \geq 1$ and $k=l=0$, and expand complex fields as follows.
\begin{equation}
\vp(z) = \Bigl[ \sum_{\substack{k=1\\l=k}}^\infty + \sum_{k=l=0} \Bigr] \; \vp_{(k,l)} F^{(p)}_{k,l}(z)
\end{equation}
Following the discussion in Section~\ref{S1Z2}, we can find the expansion of the delta function and check the completeness of our basis.


\subsection{$T^2/\Db_6$}
\label{T2D6}

The group $\pVIm$ (\ref{p6m}) is generated by the two translations $t_1$ and $t_2$ as well as the $\pi/3$-rotation $r$ and the reflection $f$. In the complex plane, we identify them with the following transformations.
\begin{align}
z &\sim z + 1\\
&\sim z + \omega \qquad \text{with} \quad \omega=\exp[i \pi/3]\\
&\sim \omega z\\
&\sim \omega z^*
\end{align}
Solving the relations in (\ref{p6m}) in $\Cb$, we find four solutions.
\begin{equation}
t_1=t_2=+, \quad r=\pm, \quad f=\pm
\end{equation}
Note that it is impossible to assign SS phases to complex fields on $T^2/\Db_6$. There is no non-trivial representation of $\pVIm$ in $\Cb$. A complex field $\vp(z)$ can posses four parities.
\begin{align}
\vp(\omega z) &= \pm \vp(z)\\
\vp(\omega z^*) &= \pm \vp(z)\\
\vp(z+1) &= \vp(z)\\
\vp(z+\omega) &= \vp(z)
\end{align}
The field can be expanded in terms of the basis $F^{(p,q)}_{k,l}(z)$ with $p,q=\pm$. Using (\ref{fkl}), (\ref{Mpi3}) and
\begin{align}
\label{T2D6sym1}
f_{k,l}(\omega z) &= f_{l,l-k}(z)\\
\label{T2D6sym2}
f_{k,l}(\omega z^*) &= f_{l,k}(z)
\end{align}
we find their explicit form.
\begin{align}
\begin{split}
F^{(p,q)}_{k,l}(z) &= c^{(p)}_{k,l} \, \Bigl( f_{k,l}(z) + p f_{l,l-k}(z) + f_{l-k,-k}(z) + p f_{-k,-l}(z) + f_{-l,k-l}(z) + p f_{k-l,k}(z)\\
&\qquad +q \bigl[ f_{l,k}(z) + p f_{l-k,l}(z) + f_{-k,l-k}(z)+ p f_{-l,-k}(z) + f_{k-l,-l}(z) + p f_{k,k-l}(z) \bigr] \Bigr)
\end{split}\\
c^{(p,q)}_{k,l} &= \Bigl[ 12 \bigl(1+\delta_{k,0}+\delta_{l,0}+\delta_{k,l}+\delta_{k,2l}+\delta_{l,2k}+6\delta_{k,0}\delta_{l,0} \bigr) \Bigr]^{-1/2}
\end{align}
The normalization constants in the above expression are derived from the orthonormality condition. In (\ref{T2D6orth}) we integrate over the torus.
\begin{equation}
\label{T2D6orth}
\int \dzz \; F^{(p,q)}_{k,l}(z) F^{(r,s)*}_{m,n}(z) = \delta_{p,q} \, \delta_{r,s} \, \delta_{k,m} \, \delta_{l,n}
\end{equation}
Note that the transformations of the indices in (\ref{T2D6sym1}) and (\ref{T2D6sym2}) are cyclic. The mass spectrum is given by (\ref{m2pi3}).
\begin{equation}
\bigl[ \pz \pzc + m_{k,l}^2\bigr] F^{(p)}_{k,l}(z) =0
\end{equation}
The following zero modes vanish.
\begin{equation}
F^{(\pm,-)}_{0,k}(z)=F^{(\pm,-)}_{k,0}(z)=F^{(\pm,-)}_{k,k}(z)=F^{(p,-p)}_{k,2k}(z)=F^{(p,-p)}_{2k,k}(z)=0 \qquad \text{with} \quad p=\pm
\end{equation}
Using (\ref{T2D6sym1}) and (\ref{T2D6sym2}), we find that the $F^{(p,q)}_{k,l}(z)$ are not independent.
\begin{gather}
F^{(p,q)}_{k,l} (\omega z) = F^{(p,q)}_{l,l-k}(z) = p \, F^{(p,q)}_{k,l}(z)\\
F^{(p,q)}_{k,l} (\omega^2 z) = F^{(p,q)}_{l-k,-k}(z) = p^2 \, F^{(p,q)}_{k,l}(z)\\
F^{(p,q)}_{k,l} (\omega z^*) = F^{(p,q)}_{l,k}(z) = q \, F^{(p,q)}_{k,l}(z)
\end{gather}
We restrict the indices $k \geq 0$ and $l \geq 2k$ and expand complex fields as follows.
\begin{equation}
\vp(z) = \sum_{\substack{k=0\\ l=2k}}^\infty \; \vp_{(k,l)} F^{(p,q)}_{k,l}(z)
\end{equation}
Following the discussion in Section~\ref{S1Z2}, we can find the expansion of the delta function and check the completeness of our basis.


\section{Conclusions}

In this paper, we present the complete classification of 1D and 2D orbifold compactifications. We derive the explicit form of the basis functions and prove their orthonormality and completeness. For each of the orbifolds, we determine all possible parities that a complex field compactified on them can possess. The classification includes the familiar orbifolds $S^1$, $S^1/\Zb_2$, $T^2$ and $T^2/\Zb_n$. We show that the compactifications $S^1/(\Zb_2 \times \Zb_2')$~\cite{Hebecker:2001wq}, $T^2/(\Zb_2 \times \Zb_2' \times \Zb_2'')$~\cite{Asaka:2002nd} and the Chiral Square~\cite{Burdman:2005sr, Dobrescu:2004zi} correspond respectively to orbifolds $S^1/\Zb_2$, $T^2/\Zb_2$ and $T^2/\Zb_4$ with non-trivial Scherk-Schwarz phases. The classification does also include a number of less familiar orbifolds with interesting new properties. For example, orbifolds such as $T^2/\Db_n$ or $T^2/\Fb_n$ posses \emph{orbifold fixed lines} that have so far not been studied in the literature. On the other hand, the Klein bottle and the real projective plane $\Rb P^2$ allow for fields with different parities, but do possess neither fixed lines nor fixed points. Hence, it is possible to define chiral matter without the difficulties that come with brane kinetic terms.

The novel \emph{purely algebraic definition} of the space groups is a central part of this paper. It allows us to determine the possible parities in a simple way. For example, the two Scherk-Schwarz phases on $T^2/\Zb_2$ are discrete but independent (\ref{p2solutions}). On $T^2/\Zb_3$ the two SS phases have to be identical (\ref{p3solutions}), and on $T^2/\Zb_6$ we cannot introduce any SS phases at all (\ref{p6solutions}). In this paper, we restrict our discussion to complex fields. In order to determine all possible parities for \emph{non-abelian} fields, it will be necessary to determine matrix solutions for the relations in (\ref{p1})-(\ref{p6m}). Note that in that case, commutators such as $[t_1,t_2]$ in definition (\ref{p4}) become relevant.

The project was guided by two principles: (i) We clearly distinguish between the definition of the orbifolds and any physics that is taking place on them, see for example (\ref{T2Z2def1})-(\ref{T2Z2def3}) and (\ref{phi1})-(\ref{phi3}). (ii) We consider the orbifolds as quotients spaces $\Rb/\Gamma$ and $\Rb^2/\Gamma$ rather than $S^1/\Gamma'$ or $T^2/\Gamma'$, cf. (\ref{1Dphilosophy}) and (\ref{2Dphilosophy}). All generators in the algebraic definitions of the space groups are equal. The translations are not singled out. Circle and torus are not special, neither are the SS phases. A parity $t_2=\pm$ in (\ref{pgsolutions}) is fundamentally not different from $r=\pm$ in (\ref{p2solutions}).

We would like to comment on the orthonormality relations in Section~\ref{S1} and Appendix~\ref{SSonT2Z2}. We define the scalar products on the functional spaces in this particular way for the following reasons: (i)~A general proof of the \emph{sum rules} requires that all non-even basis functions are orthogonal to the constant function, i.e. the even zero mode, see last paragraph in Appendix~\ref{sumrules}. We are therefore able to prove sum rules similar to (\ref{rule1})-(\ref{rule4}) for each of the orbifolds in this paper. The entire programme of~\cite{Muck:2004br, LarsPhD} (quantization, Ward identities, equivalence theorems, high energy unitarity) can easily be repeated on any of the orbifolds. (ii)~We prefer to work in a basis where any two basis functions are orthogonal to each other. Orthogonality is not restricted to the functional subspaces of equal parity. (iii)~The orthonormality relations (\ref{S1orth}) and (\ref{T2orthSS}) respect the second of the principles discussed in the paragraph above; SS phases are not singled out among the parities. If a complex field has got $n$ different parities, we integrate $n$ times over the fundamental domain of the orbifold. In Sections~\ref{T2Z3} for example, we ignore any SS phases and distinguish between three parities. We therefore integrate over the torus. In Section~\ref{S1}, complex fields can posses $j$ different SS phases, and we integrate $j$ times over the circle. Finally in Appendix~\ref{SSonT2Z2}, we find eight parities and integrate four times over the torus.

In Appendix~\ref{BKTonT2Z3}, we discuss brane kinetic terms and describe how to construct the mass eigenstate basis for compactifications on $T^2/\Zb_n$. Our derivation generalizes an earlier approach that was first discussed in~\cite{LarsPhD}.


\subsection*{Acknowledgements}
The author would like to thank Apostolos Pilaftsis for collaboration and guidance
throughout this project and in particular for pointing out the use of the
$\Db_n$ groups for some of the orbifold compactifications. We would like to thank Ignatios Antoniadis for a helpful conversation and bringing the paper~\cite{Dulat:2000xj} to our attention. The author is grateful to Arthur Hebecker for clarifying the orthonormality of the basis in~\cite{Hebecker:2001wq}.


\newpage
\begin{appendix}
\section{Sum rules on $S^1/\Zb_2$}
\label{sumrules}

In~\cite{Muck:2004br, LarsPhD} we studied the high energy unitarity of five-dimensional Yang-Mills theories compactified on an $S^1/\Zb_2$ orbifold. We found that the fundamental couplings of the 4D effective theory obey a set of non-trivial sum rules that lie at the heart of the high energy unitarity cancellations. In this appendix we present a much simplified proof of these rules and correct one of our results.

The fundamental 5D Lagrangian~(\ref{L5D}) includes an additional kinetic term that is localized at the orbifold fixed point $y=0$. This so-called \emph{brane kinetic term} (BKT) is necessary in order to renormalize operators arising from quantum corrections of the bulk fields. The dimensionful coupling $r_c$ determines the strength of the BKT and is a free parameter of the theory.
\begin{equation}
\label{L5D}
\Lag_{\rm 5D} (x,y) = - \frac{1}{4}\, \big[ 1\: +\: r_c\,\delta(y) \big] \, F^a_{MN} F^{a \; MN} + \Lag_{\rm 5D\GF} + \Lag_{\rm 5D\FP}
\end{equation}
The theory is quantized by adding the two terms $\Lag_{\rm 5D\GF}$ and $\Lag_{\rm 5D\FP}$ to the higher-dimensional Lagrangian. We work in the framework of the generalized $R_{\xi}$ gauges where the gauge-fixing functional is given by (\ref{FAaM}).
\begin{align}
\label{LGF}
\Lag_{\rm 5D\GF} &= -\, \big[1 + r_c \delta(y) \big] \, \frac{1}{2 \xi} \; \big( F[A^a_M]\big)^2\\
\label{LFP}
\Lag_{\rm 5D\FP} &= \, \big[ 1 + r_c \delta(y) \big] \, \bar{c}^a \; \frac{\delta F[A^a_M]}{\delta \theta^b} \; c^b\\
\label{FAaM}
F[A^a_M] &= \partial^\mu A^a_\mu - \xi \; \partial_5 A^a_5
\end{align}
In order to compactify the extra dimension $y$, we need to expand the higher-dimensional fields in terms of an orthonormal basis. The form of the basis functions was derived in Appendix B of~\cite{LarsPhD}.
\begin{align}
\label{fn}
f_n(y) =& \frac{N_n}{\sqrt{2^{\delta_{n,0}} \pi R} \; \cos m_n \pi R} \times \begin{cases}
\cos m_n ( y + \pi R) \; &\textrm{for} \quad -\pi R < y \leq 0\\
\cos m_n ( y - \pi R) \; &\textrm{for}\qquad 0 < y \leq \pi R
\end{cases}\\[3mm]
\label{gn}
g_n(y) =& \frac{N_n}{\sqrt{\pi R} \; \cos m_n \pi R} \times \begin{cases}
\sin m_n ( y + \pi R) \; &\textrm{for} \quad -\pi R < y < 0\\
\sin m_n ( y - \pi R) \; &\textrm{for}\qquad 0 < y \leq \pi R\\
0 &\textrm{for} \qquad y=0
\end{cases}
\end{align}
The BKT in the Lagrangian result in distortion of the standard Kaluza-Klein mass spectrum $n/R$. The spectrum $m_n$ is now determined by the transcendental equation~(\ref{BKTmn}).
\begin{align}
N_n^{-2} &= 1 + \tilde{r}_c + \pi^2 R^2 \tilde{r}_c^2 m_n^2 \qquad \textrm{with} \quad \tilde{r}_c = \frac{r_c}{2 \pi R} \geq 0\\
\label{BKTmn}
\frac{m_n r_c}{2} &= - \tan m_n \pi R
\end{align}
We can derive an expansion of the delta function in this basis by substituting an even ansatz in the defining relation~(\ref{BKTdelta}).
\begin{equation}
\label{BKTdelta}
\int_{- \pi R}^{\pi R} dy \; [1+r_c \delta(y)] \, \vp(y) \delta(y-y'; \, r_c) = \vp(y')
\end{equation}
\begin{align}
\delta(y; r_c) \; &= \; \sum_{n=0}^\infty \; f_n(0) f_n(y)\\
&= \;\begin{cases}
1/r_c \; &\textrm{for} \quad y = 0\\
0 \; &\textrm{for} \quad - \pi R < y < 0 \;\, \textrm{or} \;\, 0 < y \leq \pi R
\end{cases} \nonumber
\end{align}
The basis functions $f_n(y)$ and $g_n(y)$ are orthonormal and complete. Note that the 5D Lagrangian is proportional to a common factor $[ 1 + r_c \delta(y) ]$ which also appears in the integration measures of (\ref{BKTdelta}) and (\ref{BKTorthonormal}).
\begin{equation}
\label{BKTorthonormal}
\begin{split}
\int_{- \pi R}^{\pi R} dy \; \big[ 1 + r_c \delta(y) \big] \, f_n(y) f_m(y) &= \delta_{n,m}\\
\int_{- \pi R}^{\pi R} dy \; \big[ 1 + r_c \delta(y) \big] \, g_n(y) g_m(y) &= \delta_{n,m}
\end{split}
\end{equation}
\begin{equation}
\label{BKTcomplete}
\delta (y_1-y_2; \, r_c) = \sum_{n=0}^{\infty}\,\big[f_n (y_1) \, f_n (y_2) + g_n (y_1) \, g_n(y_2)\big]
\end{equation}
After compactifications, we find that the fundamental couplings of the effective 4D theory are proportional to coefficients such as $\Delta_{k,l,n}$, see (C.2) in~\cite{LarsPhD}. These coefficients are defined as integrals over products of even and odd basis functions.
\begin{align}
\label{Delta1}
\Delta_{k,l,n} \equiv \; &&\hspace{-1.4cm}\Delta_{k,l,n} &\equiv \sqrt{2^{\delta_{k,0}+\delta_{l,0}+\delta_{n,0}} \, 4 \pi R} \; \int_{-\pi R}^{\pi R} dy \, [1+r_c \delta(y)] f_k f_l f_n\\
\tilde{\Delta}_{k, n, l} \equiv \; &&\hspace{-1.4cm}\Delta^{k, l}_n &\equiv \sqrt{2^{\delta_{n,0}} \, 4 \pi R} \; \int_{-\pi R}^{\pi R} dy \, [1+r_c \delta(y)] g_k g_l f_n\\
\label{Delta3}
\Delta_{k,l,m,n} \equiv \; &&\hspace{-1.4cm}\Delta_{k,l,m,n} &\equiv 4 \pi R \; \sqrt{2^{\delta_{k,0}+\delta_{l,0}+\delta_{m,0}+\delta_{n,0}}} \; \int_{-\pi R}^{\pi R} dy \, [1+r_c \delta(y)] f_k f_l f_m f_n\\
\tilde{\Delta}_{m, n, k, l} \equiv \; &&\hspace{-1.4cm}\Delta^{k,l}_{m, n} &\equiv 4 \pi R \; \sqrt{2^{\delta_{m,0}+\delta_{n,0}}} \; \int_{-\pi R}^{\pi R} dy \, [1+r_c \delta(y)] g_k g_l f_m f_n\\
\label{Delta5}
&&\hspace{-1.4cm}\Delta^{k, l, m, n} &\equiv 4 \pi R \; \int_{-\pi R}^{\pi R} dy \, [1+r_c \delta(y)] g_k g_l g_m g_n
\end{align}
In this appendix, we alter our notation slightly and make use of lower and upper indices. In the new notation, the symmetry among the indices is immediately apparent. The coefficients (\ref{Delta1}) to (\ref{Delta5}) are invariant under permutations of their lower or upper indices, for example $\Delta^{k,l}_n = \Delta^{l,k}_n$.

In Appendix D of~\cite{LarsPhD}, we derive sum rules that lead to important cancellations in high energy scattering amplitudes. In what follows, we present an alternative and far less laborious derivation of these rules. Let us start with the proof of the following identity.
\begin{equation}
\label{rule1}
\sum_{j=0}^{\infty} 2^{-\delta_{j,0}} \Delta_{k,l,j} \Delta_{m,n,j} = \Delta_{k,l,m,n}
\end{equation}
Up to an overall constant, we can write the LHS of~(\ref{rule1}) as
\begin{equation*}
\sum_{j=0}^{\infty} \Big( \int_{- \pi R}^{\pi R} dy \, [1 + r_c \delta(y)] f_k(y) f_l(y) f_j(y) \Big) \Big( \int_{- \pi R}^{\pi R} dy' \, [1 + r_c \delta(y')] f_m(y') f_n(y') f_j(y') \Big) \; .
\end{equation*}
Rearranging the expression, we recognize the sum in the completeness relation~(\ref{BKTcomplete}).
\begin{equation*}
\cdots = \int_{- \pi R}^{\pi R} dy \, dy' \; [1 + r_c \delta(y)] [1 + r_c \delta(y')] f_k(y) f_l(y) f_m(y') f_n(y') \; \sum_{j=0}^{\infty} f_j(y) f_j(y')
\end{equation*}
Integrals over odd functions, such as $f_k(y) f_l(y) g_j(y)$, vanish. It is therefore only the delta function that contributes.
\begin{equation*}
\cdots = \int_{- \pi R}^{\pi R} dy \, dy' \; [1 + r_c \delta(y)] [1 + r_c \delta(y')] f_k(y) f_l(y) f_m(y') f_n(y') \; \Big( \delta(y-y'; r_c) - \sum_{j=0}^{\infty} g_j(y) g_j(y') \Big)
\end{equation*}
Using (\ref{BKTdelta}), we finally derive the RHS of (\ref{rule1}).
\begin{equation*}
\cdots = \int_{- \pi R}^{\pi R} dy \; [1 + r_c \delta(y)] f_k(y) f_l(y) f_m(y) f_n(y)
\end{equation*}
The remaining sum rules can be proven in a similar way. The argumentation relies again on the completeness of the basis.
\begin{align}
\sum_{j=0}^{\infty} 2^{-\delta_{j,0}} \Delta^{k,j}_l \Delta^{m,j}_n &= \Delta_{l,n}^{k,m}\\
\sum_{j=0}^{\infty} 2^{-\delta_{j,0}} \Delta_{k,l,j} \Delta^{m,n}_j &= \Delta^{m,n}_{k,l}\\
\label{rule4}
\sum_{j=0}^{\infty} 2^{-\delta_{j,0}} \Delta^{k, l}_j \Delta^{m,n}_j &= \Delta^{k,l,m,n} = \Delta_{k,l,m,n} + Z_{k,l,m,n}
\end{align}
It now becomes clear where the term $Z_{k,l,m,n}$ that looked so much out of place in~\cite{Muck:2004br, LarsPhD} originates from. We had not introduced a coefficient $\Delta^{k,l,m,n}$, since it did not feature in our Feynman rules. Following Appendix~C in~\cite{LarsPhD}, we can do the integration in (\ref{Delta3}) and (\ref{Delta5}) explicitly, and hence find an analytic expression for the difference of the two coefficients.
\begin{equation}
\label{Zklmn}
\begin{split}
Z_{k,l,m,n} = \; &\frac{N_k N_l N_m N_n \pi^2 R^2 \; \rct^3}{(- m_k + m_l + m_m + m_n)(m_k - m_l + m_m + m_n)}\\
&\times \frac{16(m_k m_l + m_m m_n)(m_k m_m + m_l m_n)(m_k m_n + m_l m_m)}{(m_k + m_l - m_m + m_n)(m_k + m_l + m_m - m_n)}
\end{split}
\end{equation}
Expressions (D.12) in~\cite{Muck:2004br} and (D.8) in~\cite{LarsPhD} are incorrect and need to be replaced by (\ref{Zklmn}). In the case of the two special coefficients $Y_{n,m}$ and $X_n$, we confirm our earlier results.
\begin{align}
Y_{n,m} \equiv Z_{n,n,m,m} &= 4 N_n^2 N_m^2 \pi^2 R^2 \rct^3 (m_n^2 + m_m^2)\\
X_n \equiv \; \, Z_{n,n,n,n} &= 8 N_n^4 \pi^2 R^2 \rct^3 m_n^2
\end{align}
Unlike in Appendix D of~\cite{LarsPhD}, the proof of the sum rules presented above requires no knowledge of the analytic form of the coefficients. The completeness and orthogonality of the basis functions that appear in (\ref{Delta1}) to (\ref{Delta5}) are sufficient. It is therefore not difficult to prove sum rules equivalent to (\ref{rule1})-(\ref{rule4}) for any of the 1D or 2D orbifolds discussed in this paper. The entire formalism of Ward and Slavnov-Taylor identities introduced in~\cite{Muck:2004br, LarsPhD} can readily be used in all of these cases.

The last step of the proof of the sum rule~(\ref{rule1}) relies on the fact that integrals over odd functions do not contribute. If we want to prove similar rules on other orbifolds, we need to require that integrals over any non-even basis function vanish. In other words, basis function of different parities (including different Scherk-Schwarz parities) should be orthogonal to each other. It is for this reason that we prefer to work with orthonormalities~(\ref{S1orth}) and~(\ref{T2orthSS}) in this paper.


\section{Scherk-Schwarz mechanism on $T^2/\Zb_2$}
\label{SSonT2Z2}

From our discussion of one-dimensional orbifolds we know that the mass spectra and exponentials possess a dependence on the Scherk-Schwarz phases, see (\ref{fk}) and (\ref{S1wave}). In order to keep our notation simple and the number of indices to a minimum, we ignored SS phases in our study of 2D orbifolds in Sections~\ref{T2} to~\ref{T2D6}. In what follows, we will demonstrate that the bases can easily be modified in order to accommodate SS phases.

Let us consider the example of $T^2/\Zb_2$. The construction of the basis of this orbifold relies on the basis of the torus $T^2$. Let us therefore retrace the steps of Section~\ref{T2}, but now allow for SS phases.
\begin{align}
\vp(z+1) &= \exp[i 2 \pi \rho_1] \, \vp(z)\\
\vp(z+\omega) &= \exp[i 2 \pi \rho_2] \, \vp(z) \qquad \text{with} \quad \rho_{1,2} \in [0,1) \subset \Qb
\end{align}
Complex fields $\vp(z)$ are no longer periodic but acquire SS phases $p_{1,2}=\exp[i 2 \pi \rho_{1,2}]$. The fields can be expanded in terms of exponentials $f^{(p_1,p_2)}_{k,l}(z)$.
\begin{align}
\label{fklSS}
f^{(p_1,p_2)}_{k,l}(z) &= \exp \bigl[ i 2 \pi ([k+\rho_1] y_5 + [l+\rho_2] y_6) \bigr]\\
\label{fklSSII}
&=\exp \bigl[i 2 \operatorname{Re}(M^{(p_1,p_2)}_{k,l} z)\bigr]\\
M^{(p_1,p_2)}_{k,l} &= 2 \pi \; \frac{l+\rho_2-\omega^* (k+\rho_1)}{\omega-\omega^*}
\end{align}
\begin{equation}
\vp(z) = \sum_{k,l=-\infty}^{\infty} \; \vp_{(k,l)} f^{(p_1,p_2)}_{k,l}(z)
\end{equation}
In the following, let $j_1, j_2 \in \Nb$ the smallest integers such that $p^{j_1}=p^{j_2}=1$. The basis functions (\ref{fklSS}) are completely orthonormal in the following sense.
\begin{align}
\frac{1}{j_1 j_2 r \, \sin \theta}\int_0^{j_1} dy_5 \; \int_0^{j_2}dy_6 \; f^{(p_1,p_2)}_{k,l}(y_5,y_6) f^{(q_1,q_2)*}_{m,n}(y_5,y_6) &= \delta_{p_1,q_1} \delta_{p_2,q_2} \delta_{k,l} \delta_{m,n} \nonumber\\
\label{T2orthSS}
\int \frac{i \; dz dz^*}{j_1 j_2 \; 2(\operatorname{Im} \omega)^2} \; f^{(p_1,p_2)}_{k,l}(z) f^{(q_1,q_2)*}_{m,n}(z) &= \delta_{p_1,q_1} \delta_{p_2,q_2} \delta_{k,l} \delta_{m,n}
\end{align}
The above expression generalizes the one-dimensional case (\ref{S1orth}). It can be checked in cartesian coordinates $(y_5,y_6)$, where we integrate $j_1 j_2$ times over the fundamental domain of the torus. In this appendix, we will again prefer the complex notation (\ref{T2orthSS}). -- As expected, we find that the mass spectrum $m^{(p_1,p_2)}_{k,l}$ depends on the two SS phases $p_1$ and $p_2$.
\begin{equation}
\begin{split}
\pz f^{(p_1,p_2)}_{k,l}(z) &= i M^{(p_1,p_2)}_{k,l} f^{(p_1,p_2)}_{k,l}(z)\\
\pzc f^{(p_1,p_2)}_{k,l}(z) &= i M^{(p_1,p_2)*}_{k,l} f^{(p_1,p_2)}_{k,l}(z)
\end{split}
\end{equation}
\begin{equation}
\bigl[ \pz \pzc + m^{(p_1,p_2)2}_{k,l}\bigr] f^{(p_1,p_2)}_{k,l}(z) = 0 \qquad \text{with} \quad m^{(p_1,p_2)2}_{k,l} \equiv M_{k,l}^{(p_1,p_2)} M_{k,l}^{(p_1,p_2)*}
\end{equation}
There exists a relation between indices and arguments of the exponentials that will prove to be useful later on.
\begin{equation}
\label{T2symSS}
f^{(p,q)}_{k,l}(-z)=f^{(p^*,q^*)}_{-k,-l}(z)
\end{equation}
Let us now turn to $T^2/\Zb_2$. As derived in (\ref{p2solutions}), a complex field $\vp(z)$ on this orbifold can posses eight different parities $(p,q,s)$.
\begin{align}
\label{vpI}
\vp(z+1) &= p \; \vp(z)\\
\vp(z+i) &= q \; \vp(z)\\
\label{vpIII}
\vp(- z) &= s \; \vp(z)
\end{align}
Since $p,q,s=\pm$, we find $j_1=j_2=2$. There are two free parameters that determine the shape of $T^2/\Zb_2$, see Table~2. In this appendix, we choose $\omega = r \exp[i \theta] = i$. The complex field $\vp(z)$ can be expanded in terms of functions $F^{(p,q,s)}_{k,l}(z)$. Using (\ref{fklSSII}) and (\ref{T2symSS}), we find their explicit form.
\begin{equation}
\label{FklSS}
F^{(p,q,\pm)}_{k,l}(z) = \sqrt{2^{-1-\delta_{k,0} \delta_{l,0}}} \Bigl[ f^{(p,q)}_{k,l}(z) \pm f^{(p,q)}_{-k,-l}(z) \Bigr]
\end{equation}
\begin{equation}
M^{(p,q)}_{k,l} = \pi \begin{cases} k - i l \qquad &\text{for} \quad (p,q)=(+,+)\\
k - i l+ 1/2 \qquad &\text{for} \quad (p,q)=(-,+)\\
k - i l- i/2 \qquad &\text{for} \quad (p,q)=(+,-)\\
k - i l+ 1/2 - i/2 \qquad &\text{for} \quad (p,q)=(-,-) \end{cases}
\end{equation}
The normalization constants in (\ref{FklSS}) are derived from the orthonormality condition below. We integrate eight times over the fundamental domain of $T^2/\Zb_2$, i.e. $y_{5,6}=0 \to 2$ in Figure~3(a).
\begin{equation}
\label{T2Z2orthSS}
\int \frac{i}{8} \; dz dz^* \; F^{(p_1,p_2,s)}_{k,l}(z) F^{(q_1,q_2,t)*}_{m,n}(z) = \delta_{p_1,q_1} \, \delta_{p_2,q_2} \, \delta_{s,t} \, \delta_{k,m} \, \delta_{l,n}
\end{equation}
Following the discussion of Section~\ref{T2Z2}, we find again that the basis functions $F^{(p,q,s)}_{k,l}(z)$ are not independent. Consequently, we restrict the indices in the expansion.
\begin{equation}
\vp(z) = \sum_{\substack{k=-\infty\\ l=1 \; \text{for} \; k < 0\\ l=0 \; \text{for} \; k \geq 0}}^\infty \vp_{(p,q,s)} F^{(p,q,s)}_{k,l}(z)
\end{equation}
The expressions below summarize the properties of the basis. 
\begin{align}
F^{(\pm,q,s)}_{k,l}(z+1) &= \pm F^{(\pm,q,s)}_{k,l}(z)\\
F^{(p,\pm,s)}_{k,l}(z+i) &= \pm F^{(p,\pm,s)}_{k,l}(z)\\
F^{(p,q,\pm)}_{k,l}(-z) &= \pm F^{(p,q,\pm)}_{k,l}(z)\\
\pz F^{(p,q,\pm)}_{k,l}(z) &= i M^{(p,q)}_{k,l} \; F^{(p,q,\mp)}_{k,l}(z)\\
\pzc F^{(p,q,\pm)}_{k,l}(z) &= i M^{(p,q)*}_{k,l} \; F^{(p,q,\mp)}_{k,l}(z)\\
\bigl[ \pz \pzc + m_{k,l}^{(p,q)2}\bigr] F^{(p,q,s)}_{k,l}(z) &=0
\end{align}
The mass spectrum $m^{(p,q)}_{k,l}$ depends on the two SS phases, $p$ and $q$.
\begin{equation}
m^{(p,q)2}_{k,l} = \pi^2 \begin{cases} k^2 + l^2 \qquad &\text{for} \quad (p,q)=(+,+)\\
k^2 + (l+1/2)^2 \qquad &\text{for} \quad (p,q)=(+,-)\\
(k+1/2)^2 + l^2 \qquad &\text{for} \quad (p,q)=(-,+)\\
(k+1/2)^2 + (l+1/2)^2 \qquad &\text{for} \quad (p,q)=(-,-) \end{cases}
\end{equation}
As in Section~\ref{S1Z2}, we can derive an expansion of the delta function and check the completeness of our basis. The completeness together with the orthonormality~(\ref{T2Z2orthSS}) allows us to derive sum rules similar to the ones of Appendix~\ref{sumrules}.
\begin{gather}
\delta(z) = \sum_{\substack{k=-\infty\\ l=1 \; \text{for} \; k < 0\\ l=0 \; \text{for} \; k \geq 0}}^\infty F^{(+,+,+)*}_{k,l}(0) F^{(+,+,+)}_{k,l}(z)\\
\delta (z_1-z_2) = \sum_{p,q,s=\pm}\sum_{\substack{k=-\infty\\ l=1 \; \text{for} \; k < 0\\ l=0 \; \text{for} \; k \geq 0}}^\infty \, F^{(p,q,r)}_{k,l} (z_1) \, F^{(p,q,r)*}_{k,l} (z_2)
\end{gather}
In this final paragraph, we would like to describe how $T^2/(\Zb_2 \times \Zb_2' \times \Zb_2'')$, as introduced by Asaka et al. in~\cite{Asaka:2002nd}, fits in our classification of orbifolds. Fields of parity $(p,q,s)$ on $T^2/\Zb_2$ with radii $R_1$ and $R_2$ correspond to parities $(s,sp,sq)=(\eta_I,\eta_{PS},\eta_{GG})$ on $T^2/(\Zb_2 \times \Zb_2' \times \Zb_2'')$ with radii $2R_1$ and $2R_2$, see definitions (3)-(5) in \cite{Asaka:2002nd}. -- The arguments should be familiar from our discussion of $S^1/(\Zb_2 \times \Zb_2')$ in Section~\ref{S1Z2}. The choice of generators of the space group~$\pII$ is not unique. In (\ref{vpI})-(\ref{vpIII}), we assign parities to $t_1$, $t_2$ and $r$. Asaka et al. choose to work with the generators $r$, $t_1 r$ and $t_2 r$ instead.


\section{Brane kinetic terms on $T^2/\Zb_3$}
\label{BKTonT2Z3}

In Appendix~B of~\cite{LarsPhD} we derived the basis functions for a 5D quantum field theory compactified on $S^1/\Zb_2$ with a brane kinetic term (BKT) at $y=0$. In this appendix, we would like to demonstrate that the same approach can be used to find the basis functions for the compactification of 6D theories on $T^2/\Zb_n$ with BKTs at the orbifold fixed points.

Consider a 5D theory compactified on $S^1/\Zb_2$ \emph{without} BKTs. The basis functions fulfil a number of relations, such as orthonormality, wave equations and parity relations. Modifying merely the integration measure and the spectrum
\begin{align}
\label{IM5D}
dy \quad &\rightarrow \quad dy \, \bigl[ 1 + r_c \delta (y) \bigr]\\
\label{m5D}
n/R \quad &\rightarrow \quad m_n
\end{align}
we find equations (B.1), (B.2), (B.5) and (B.7) in~\cite{LarsPhD}. The relations fulfilled by the basis with and the basis without BKTs are very similar. The new integration measure takes the BKT at $y=0$ into account, whereas $m_n$ reflects the new spectrum. Substituting an ansatz into these relations, we find the basis functions (\ref{fn}) and (\ref{gn}).

Let us now consider the example of a 6D theory compactified on $T^2/\Zb_3$ with BKTs at the orbifold fixed points $z=0$, $z=i/\sqrt{3}$ and $z=1/2+i/(2 \sqrt{3})$, see Figure 6. Following closely the above approach, we will outline the derivation of the mass eigenmode functions $G^{(p)}_{k,l}(z)$ of the theory. We demand the basis functions to obey the following relations.
\begin{align}
\label{R1}
G^{(p)}_{k,l}(\omega z) &= p \, G^{(p)}_{k,l}(z)\\
\label{R2}
\pz G^{(p)}_{k,l}(z) &= i M_{k,l} G^{(\omega^* p)}_{k,l}(z)\\
\label{R3}
\bigl[ \pz \pzc + m_{k,l}^2 \bigr] G^{(p)}_{k,l}(z) &= 0 \qquad \text{with} \quad m_{k,l}^2 \equiv M_{k,l} M^*_{k,l}\\
\label{R4}
\int \dBKT \, G^{(p)}_{k,l}(z) G^{(q)*}_{m,n}(z) &= \delta_{p,q} \delta_{k,m} \delta_{l,n}
\end{align}
These relations are almost identical to the ones encountered in Section~\ref{T2Z3}. The integration measure in the orthonormality relation differs by a multiplicative factor which reflects the presence of the BKTs.
\begin{equation}
\dBKT \equiv \dzz \; \Bigl[ 1 + r_c \delta^*\bigl(z\bigr) + r_c \delta^*\bigl(z-\frac{i}{\sqrt{3}}\bigr) + r_c \delta^* \bigl(z-\frac{1}{2}-\frac{i}{2 \sqrt{3}}\bigr) \Bigr]
\end{equation}
The $G^{(p)}_{k,l}(z)$ are linear combinations of (\ref{T2Z3F}) which are themselves superpositions of (\ref{fkl}). The coefficients $M_{k,l}$ are replaced by 
\begin{equation}
M_{k,l}=\frac{m_l - \omega^* \tilde{m}_k}{\omega - \omega^*}
\end{equation}
with $\omega=\exp [i 2 \pi/3]$. It is no longer necessary to demand $l$ and $k$ in (\ref{M}) do be integers, since the $G^{(p)}_{k,l}(z)$ are periodic by construction.

Complex fields on $S^1/\Zb_2$ can either be even or odd\footnote{In this appendix, we ignore any Scherk-Schwarz phases. All fields are invariant under translations, that is under transformations $t \in \Db_\infty$ or $t_{1,2} \in \pIII$.}. In the orthonormality relation in~\cite{LarsPhD} we integrate twice over the fundamental domain of $S^1/\Zb_2$, i.e. over the fundamental domain of the circle. The two regions $-\pi R < y \leq 0$ and $0 < y \leq \pi R$ are related by the transformation $y \to -y$. It corresponds to the group element $r \in \Db_\infty$. -- Fields on $T^2/\Zb_3$ can possess three parities and we consequently integrate three times over the fundamental domain of the orbifold, i.e. over the torus. We divide the torus into three regions as indicated in Figure 6.
\begin{figure}[!tbp]
  \begin{center}
  \begin{minipage}[t]{0.75\textwidth}
  \centering
  \begin{minipage}[t]{0.45\textwidth}
    \parbox[b]{\textwidth}{
      \centering
      \includegraphics{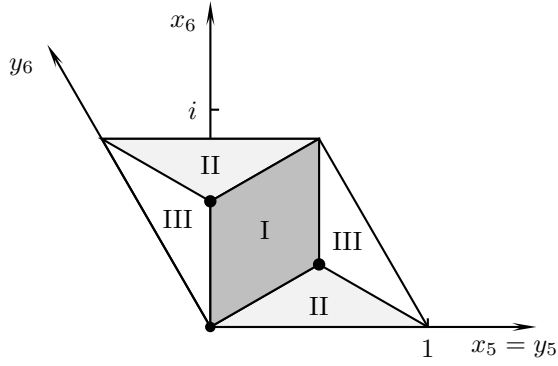}}
  \end{minipage}
  \caption{We divide the fundamental domain of the torus into three regions. Region~$\I$: fundamental domain of $T^2/\Zb_3$. Region~$\II$: $t_1 r, \, t_2 t_1 r \in \pIII$ acting on region~$\I$. Region~$\III$: $t_2 r^2, \, t_1 t_2 r^2 \in \pIII$ acting on region~$\I$.}
  \label{figT2Z3BKT}
  \end{minipage}
  \end{center}
\end{figure}

Let us make the following ansatz: In region $\I$, the even mass eigenmode function $G^{(+)}_{k,l}(z)$ is a linear combination of the $F^{(p)}_{k,l}(z)$. Due to (\ref{T2Z3wave}), $G^{(+)}_{k,l}(z)$ obeys therefore the wave equation~(\ref{R3}). From the parity relation (\ref{R1}) we can deduce the form of $G^{(+)}_{k,l}(z)$ in regions $\II$ and $\III$. Using (\ref{R2}), we find subsequently the form of the other two parities, $G^{(\omega)}_{k,l}(z)$ and $G^{(\omega^2)}_{k,l}(z)$.
\begin{align}
\label{A1}
G^{(+)}_{k,l}(z) &= 
\begin{cases}
A \, F^{(+)}_{k,l}(z) + B \, F^{(\omega)}_{k,l}(z) + C \, F^{(\omega^2)}_{k,l}(z) \qquad &\; \; \text{for} \quad z \in \I\\
A \, F^{(+)}_{k,l}(z) + B \omega^2 \, F^{(\omega)}_{k,l}(z) + C \omega \, F^{(\omega^2)}_{k,l}(z) \qquad &\; \; \text{for} \quad z \in \II\\
A \, F^{(+)}_{k,l}(z) + B \omega \, F^{(\omega)}_{k,l}(z) + C \omega^2 \, F^{(\omega^2)}_{k,l}(z) \qquad &\; \; \text{for} \quad z \in \III
\end{cases}\\
G^{(\omega)}_{k,l}(z) &= 
\begin{cases}
C \, F^{(+)}_{k,l}(z) + A \, F^{(\omega)}_{k,l}(z) + B \, F^{(\omega^2)}_{k,l}(z) \qquad &\; \; \text{for} \quad z \in \I\\
C \omega \, F^{(+)}_{k,l}(z) + A \, F^{(\omega)}_{k,l}(z) + B \omega^2 \, F^{(\omega^2)}_{k,l}(z) \qquad &\; \; \text{for} \quad z \in \II\\
C \omega^2 \, F^{(+)}_{k,l}(z) + A \, F^{(\omega)}_{k,l}(z) + B \omega \, F^{(\omega^2)}_{k,l}(z) \qquad &\; \; \text{for} \quad z \in \III
\end{cases}\\
\label{A3}
G^{(\omega^2)}_{k,l}(z) &= 
\begin{cases}
B \, F^{(+)}_{k,l}(z) + C \, F^{(\omega)}_{k,l}(z) + A \, F^{(\omega^2)}_{k,l}(z) \qquad &\text{for} \quad z \in \I\\
B \omega^2 \, F^{(+)}_{k,l}(z) + C \omega \, F^{(\omega)}_{k,l}(z) + A \omega \, F^{(\omega^2)}_{k,l}(z) \qquad &\text{for} \quad z \in \II\\
B \omega \, F^{(+)}_{k,l}(z) + C \omega^2 \, F^{(\omega)}_{k,l}(z) + A \, F^{(\omega^2)}_{k,l}(z) \qquad &\text{for} \quad z \in \III
\end{cases}
\end{align}
Substituting the expressions above into the four orthonormality relations
\begin{align}
\label{O1}
\int \dBKT \, G^{(p)}_{k,l}(z) G^{(p)*}_{k,l}(z) &= 1 \qquad \text{with} \quad p \in \{+,\omega,\omega^2\}\\
\label{O2}
\int \dBKT \, G^{(+)}_{k,l}(z) &= \delta_{k,0} \delta_{l,0}
\end{align}
we are able to determine the four complex unknowns of the ansatz, $A, B, C \in \Cb$ and $m_l, \tilde{m}_k \in \Rb$. Note that we follow indeed very closely the derivation on $S^1/\Zb_2$. The ansatz~(\ref{A1})-(\ref{A3}) and the relations~(\ref{O1})-(\ref{O2}) correspond to (B.8) to~(B.10) in~\cite{LarsPhD}.

It is not difficult to modify the above approach and derive the mass eigenstate bases for BKT theories compactified on $T^2/\Zb_n$. The $n+1$ unknowns of the ansatz are fixed by $n+1$ orthonormality relations. Since the bases are orthonormal and complete, we can derive sum rules similar to the ones discussed in Appendix~\ref{sumrules}.

\end{appendix}


\end{document}